\documentclass[11pt]{article}
\pdfoutput=1
\usepackage{putex}
\usepackage{amsmath,hyperref,amssymb,amsfonts,bm,amscd,slashed}
\usepackage{cite}
\usepackage{graphicx,subcaption}
\usepackage[all]{xy}
\usepackage{appendix}
\usepackage{array}
\usepackage{url}
\usepackage{float}
\usepackage{mathtools}
\usepackage{mathrsfs}
\usepackage{bbm}

\numberwithin{equation}{section}

\newcommand {\be} {\begin {equation}}
\newcommand {\ee} {\end {equation}}

\newcommand {\bes} {\begin {equation*}}
\newcommand {\ees} {\end {equation*}}

\newcommand{\Z}{\mathbb{Z}}

\newcommand{\R}{\mathbb{R}}
\newcommand{\C}{\mathbb{C}}

\def\la{\langle}
\def\ra{\rangle}
\def\pa{\partial}

\def\ch{{\rm ch}}
\def\sh{{\rm sh}}

\newcommand{\beq}{\begin{equation}}
\newcommand{\eeq}{\end{equation}}

\def\<{\langle}
\def\>{\rangle}

\newcommand{\bC}{\ensuremath{\mathbb{C}}}

\newcommand{\bZ}{\ensuremath{\mathbb{Z}}}

\newcommand{\cC}{\ensuremath{\mathcal{C}}}
\newcommand{\cD}{\ensuremath{\mathcal{D}}}

\newcommand{\cI}{\ensuremath{\mathcal{I}}}
\newcommand{\cJ}{\ensuremath{\mathcal{J}}}

\newcommand{\cL}{\ensuremath{\mathcal{L}}}
\newcommand{\cM}{\ensuremath{\mathcal{M}}}
\newcommand{\cN}{\ensuremath{\mathcal{N}}}
\newcommand{\cO}{\ensuremath{\mathcal{O}}}

\newcommand{\cR}{\ensuremath{\mathcal{R}}}
\newcommand{\cS}{\ensuremath{\mathcal{S}}}
\newcommand{\cT}{\ensuremath{\mathcal{T}}}
\newcommand{\cU}{\ensuremath{\mathcal{U}}}

\newcommand{\cW}{\ensuremath{\mathcal{W}}}

\newcommand{\dd}{\mathrm{d}}

\numberwithin{equation}{section}
\DeclareMathOperator{\trace}{Tr}

\usepackage{amsthm}

\theoremstyle{plain}

\theoremstyle{definition}

\def\ie{\begin{equation}\begin{aligned}}
\def\fe{\end{aligned}\end{equation}}

\newcommand{\B}{{\beta}}

\newcommand{\D}{{\delta}}
\newcommand{\mf}{\mathfrak }

\usepackage{tocloft}

\begin{document}
\preprint{CALT-TH 2019-041\\ PUPT-2602}

\institution{SCGP}{Simons Center for Geometry and Physics,	Stony Brook University,\cr Stony Brook NY 11794 USA}
\institution{Caltech}{Walter Burke Institute for Theoretical Physics, California Institute of Technology, \cr Pasadena, CA 91125, USA}
	\institution{PU}{Joseph Henry Laboratories, Princeton University, Princeton, NJ 08544, USA}
	\institution{CMSA}{Center of Mathematical Sciences and Applications, Harvard University, Cambridge, MA 02138, USA}
	\institution{HU}{Jefferson Physical Laboratory, Harvard University,
		Cambridge, MA 02138, USA}

\title{4d/2d $\rightarrow $ 3d/1d \\ A song of protected operator algebras}
\authors{Mykola Dedushenko\worksat{\SCGP,\Caltech} and Yifan Wang\worksat{\PU,\CMSA,\HU}}

\abstract{
	Superconformal field theories (SCFT) are known to possess solvable yet nontrivial sectors in their full operator algebras. Two prime examples are the chiral algebra sector on a two dimensional plane in   four  dimensional $\cN=2$ SCFTs, and the topological quantum mechanics (TQM) sector on a line in three dimensional $\cN=4$ SCFTs.  Under Weyl transformation, they respectively map to operator algebras on a great torus in $S^1\times S^3$ and a great circle in $S^3$, and are naturally related by reduction along the $S^1$ factor, which amounts to taking the Cardy (high-temperature) limit of the four dimensional theory on $S^1\times S^3$.
	We elaborate on this relation by explicit examples that involve both Lagrangian and non-Lagrangian theories in four dimensions, where the chiral algebra sector is generally described by a certain W-algebra, while the three dimensional descendant SCFT always has a (mirror) Lagrangian description. 
	By taking into account a subtle R-symmetry mixing, we provide explicit dictionaries between selected operator product expansion (OPE) data in the four and three dimensional SCFTs, which we verify in the examples using recent localization results in four and three dimensions. Our methods thus provide nontrivial support for various chiral algebra proposals in the literature.
	Along the way, we also identify three dimensional mirrors for Argyres-Douglas theories of type $(A_1, D_{2n+1})$ reduced on $S^1$, and find more evidence for earlier proposals in the case of $(A_1, A_{2n-2})$,
	%	we also provide nontrivial evidence for unfamiliar three dimensional (mirror) theories from Argyres-Douglas theories of types $(A_1, A_{2n-2})$ and $(A_1, D_{2n+1})$ reduced on $S^1$, 
%	we also identify  previously unknown three dimensional (mirror) theories from Argyres-Douglas theories of types $(A_1, A_{2n-2})$ and $(A_1, D_{2n+1})$ reduced on $S^1$, 
	which both realize certain superconformal boundary conditions for the four dimensional $\cN=4$ super-Yang-Mills. This is a companion paper to arXiv:1911.05741.
}
\date{}

\maketitle
\renewcommand{\baselinestretch}{1.2}\normalsize
\tableofcontents
\renewcommand{\baselinestretch}{1.2}\normalsize

\setlength{\unitlength}{2mm}

\newpage

\section{Introduction}
This paper reports on a recent progress in understanding two similar constructions in supersymmetric field theories with eight supercharges via exploring examples and applications. Our focus is on the relation between vertex algebras in 4d $\cN=2$ superconformal field theories (SCFTs) \cite{Beem:2013sza} and 1d TQFTs, or topological quantum mechanics (TQM), in 3d $\cN=4$ field theories \cite{Chester:2014mea,Beem:2016cbd,Dedushenko:2016jxl}. These algebraic structures are two most studied examples in the family of constructions that identify lower-dimensional field theories in the cohomology of higher-dimensional field theories with extended supersymmetry \cite{Beem:2013sza,Chester:2014mea,Beem:2016cbd,Dedushenko:2016jxl,Beem:2014kka,Dedushenko:2017avn,Dedushenko:2018icp,Mezei:2018url}.

Recent months have seen exciting new developments on the three-dimensional side: \cite{ERS1} initiate the mathematical study of ``short'' star products that appear in 3d, and describe certain classification results; \cite{Gaiotto:2019mmf} propose and test in a large set of examples an intriguing IR formula for the twisted trace capturing the TQM, which is a direct analog of the IR formula for the Schur index of 4d $\cN=2$ theories \cite{Cordova:2015nma,Cordova:2016uwk}; in \cite{Chang:2019dzt} the authors study TQM in cases when it describes quantization of the minimal nilpotent orbit of a complex simple Lie algebra and carry out the bootstrap analysis for 3d SCFTs realizing the TQM. In a companion paper \cite{Dedushenko:2019mzv}, a connection between the VOA in 4d and the TQM in 3d was explained on general grounds, and in particular relation to the non-commutative Zhu algebra \cite{Zhu} of a VOA was spelled out. 

The goal of this paper is threefold: first, we expand \cite{Dedushenko:2019mzv} by numerous examples and applications, verifying and providing alternative arguments for the statements made there; second, we apply it to propose and test previously unknown 3d mirror duals for a class of Argyres-Douglas theories \cite{Argyres:1995jj,Argyres:1995xn}; finally, the computations we do serve as consistency checks for the statements previously made in the literature that aims to identify VOAs of certain 4d $\cN=2$ SCFTs \cite{Beem:2013sza,Beem:2014rza,Lemos:2014lua,Cordova:2015nma,Nishinaka:2016hbw,Creutzig:2017qyf,Beem:2017ooy,Song:2017oew,Choi:2017nur,Creutzig:2018lbc,Arakawa:2018egx,Beem:2019tfp,Beem:2019snk}, though we do not consider all examples proposed there. Some other recent literature on the subject also includes  \cite{Buican:2015tda,Lemos:2015orc,Buican:2016arp,Arakawa:2016hkg,Bonetti:2016nma,Song:2016yfd,Fredrickson:2017yka,Buican:2017fiq,Neitzke:2017cxz,Bonetti:2018fqz,Beem:2018duj,Costello:2018zrm,Buican:2019huq,Xie:2019yds,Xie:2019zlb,Fluder:2019dpf,Bianchi:2019sxz}.

We consider the high temperature limits of 4d $\cN=2$ SCFTs placed on $S^3\times S^1$, which is the proper choice of background to address the question of dimensional reduction of the VOA construction \cite{Dedushenko:2019mzv}.\footnote{It is conceivable that the version of $\Omega$-background \cite{Nekrasov:2002qd,Nekrasov:2003rj,Nekrasov:2010ka} constructed in \cite{Oh:2019bgz,Jeong:2019pzg} can also be used, at least for answering some questions. The Omega background was used to quantize Higgs and Coulomb branches in \cite{Yagi:2014toa,Bullimore:2015lsa,Bullimore:2016hdc}, however the relation to TQM discussed here must involve some highly non-trivial change of basis.} We always assume the divergent behavior to be controlled by the Cardy-Di Pietro-Komargodski formula \cite{CARDY1986186,DiPietro:2014bca,Buican:2015ina,Ardehali:2015bla,Chang:2019uag}, and that after subtraction of the appropriate 3d supergravity counter-terms\footnote{See also related classification of counter-terms in 4d new minimal supergravity \cite{Assel:2014tba}.}   \cite{Closset:2012vp,Closset:2012vg,Closset:2012ru}, we are left with the unambiguous finite piece interpreted as the $S^3$ partition function\footnote{Reduction of indices to sphere partition functions were previously considered in \cite{Dolan:2011rp,Gadde:2011ia,Imamura:2011uw,Aharony:2013dha,Buican:2015hsa}.} of the 3d $\cN=4$ theory \cite{Dedushenko:2019mzv},\footnote{There can be scheme dependence at the subleading order $\cO(\B)$ (where $\B$ is the size of $S^1$), as explained in the recent paper \cite{Closset:2019ucb}.} which works well when the 4d Weyl anomalies obey $c_{\rm 4d}>a_{\rm 4d}$ \cite{Ardehali:2015bla,DiPietro:2016ond}.

First we study theories whose chiral algebra is the affine VOA for a simple Lie algebra $\mathfrak{g}$, in which case the corresponding TQM is described by a short star product on the filtered quantization of the minimal nilpotent orbit of $\mathfrak{g}$. The quantum algebra in this case is the quotient of $\cU(\mathfrak{g})$, the universal enveloping algebra of $\mathfrak{g}$, by the Joseph ideal. Then we consider more involved examples of VOAs given by W-algebras. Due to the relation to the Zhu algebra \cite{Zhu} found in \cite{Dedushenko:2019mzv}, it is clear that the corresponding TQM should be described by (a quotient of) the finite W-algebra \cite{DSK}, though we do not probe this structure in full generality and mostly focus on various subalgebras. In this case, we first consider a Lagrangian example of $SU(N)$ SQCD, mostly for $N=3$, and then move to the $(A_1, A_{2n+1})$ and  $(A_1, D_{2n})$ Argyres-Douglas theories.\footnote{See also \cite{Benvenuti:2017lle,Benvenuti:2017kud, Benvenuti:2017bpg} where the $S^1$ reduction is studied using the $\cN=1$ Lagrangian of the Argyres-Douglas theories \cite{Maruyoshi:2016tqk,Maruyoshi:2016aim}.} In all these examples, various mixing phenomena between abelian symmetries play central roles. One of them is the mixing between the $U(1)_r$ conformal R-symmetry in 4d and the Coulomb branch (or topological) global symmetries that emerge in 3d \cite{Buican:2015hsa}, which is expected to happen  in 4d $\cN=2$ SCFTs reduced on the circle generically, and in particular plays an important role for Argyres-Douglas theories, as we also mention in the next paragraph.

We also use our framework to propose and test previously unknown 3d mirrors for $(A_1, D_{2n+1})$ Argyres-Douglas theories (as well as test closely related $(A_1, A_{2n-2})$ theories, whose free mirrors were proposed in \cite{Benvenuti:2018bav}). In the process, we also employ a few other, non-VOA, techniques to verify our claims, such as the Coulomb branch index on lens spaces \cite{Fredrickson:2017yka}, and the constructions of these theories using the 4d $\cN=4$, or maximal, super Yang-Mills (MSYM) dimensionally reduced on the interval, relying on \cite{Gaiotto:2008sa,Gaiotto:2008ak,Chacaltana:2012zy,Benini:2010uu}. As mentioned in the previous paragraph, for $S^1$ reductions of generic 4d $\cN=2$ SCFTs, the relation between 4d conformal $U(1)_r$ and the 3d conformal $SU(2)_C$ R-symmetries involves mixing with the Coulomb branch symmetries noticed in \cite{Buican:2015hsa}. For $(A_1, D_{2n+1})$ theories, in particular, this is crucial for proper identification of their 3d reductions, since mixing generates imaginary Fayet-Iliopoulos (FI) parameters in 3d \cite{Festuccia:2011ws} that deform the TQM relations.

In the end, we find plenty of evidence that the 3d mirror of a $(A_1, D_{2n+1})$ theory can be described by $n$ decoupled sectors, one of which is the SQED$_2$, also known as the $T[SU(2)]$ SCFT, and the others are simply free 3d hypermultiplets. Correspondingly, the direct dimensional reduction is given by the $T[SU(2)]$ theory (which is self-mirror) and $n-1$ decoupled free twisted hypermultiplets. Each twisted hypermultiplet is of course well-known to be dual to SQED$_1$. A closely related claim, which we also verify (and explain its relation to $(A_1, D_{2n+1})$) is that the $(A_1, A_{2n-2})$ theory simply reduces to $n-1$ free twisted hypermultiplets, \emph{i.e.} its 3d mirror is a set of $n-1$ free hypermultiplets. This latter example is slightly outside the main scope of this paper, as the VOA of this theory is $C_2$-cofinite and is completely lifted in the 3d limit. Nonetheless, the 3d limit of the Schur index still carries   non-trivial information that allows to check our claim.

The structure of this paper is as follows. We start in Section \ref{sec:affineVOA} with the sample analysis of the torus correlation functions for the affine VOA, and their high-temperature (or small complex structure $\tau\to+0i$) limit. We call it the sample analysis because its basic features keep showing up in later examples when we either study particular affine VOAs, or look at the affine subalgebras of more complicated VOAs. In Section \ref{sec:MinimalNilp} we focus on the Deligne-Cvitanovi\'c (DC) exceptional series of SCFTs \cite{Beem:2017ooy}, as well as $(A_1, D_{2n+1})$ Argyres-Douglas theories, as these possess affine VOAs as their chiral algebras, minimal nilpotent orbits as their Higgs branches, and the TQM gives quantization of the latter. In Section \ref{sec:checkFTH} we perform an extensive analysis of $(A_1, D_{2n+1})$ and $(A_1, A_{2n-2})$ theories, proposing and testing their 3d mirrors. We continue in section \ref{sec:LagrangianEx} with Lagrangian examples, often focusing on technical details. In Section \ref{sec:ADMore} we focus on $(A_1, A_{2n-1})$ and $(A_1, D_{2n+2})$ theories, which are related by Higgsing like the $(A_1, D_{2n+1})$ and $(A_1, A_{2n-2})$ in earlier Sections. Some of the more cumbersome computations are described in Appendices.

\emph{Note added: } during the final stage of preparation of this article, the paper \cite{Pan:2019shz} appeared, which has some overlap with our results.

\section{Sample analysis: affine VOA}\label{sec:affineVOA}
Let us start with the high-temperature limit of vacuum torus correlators for affine VOAs at the non-critical level. This class of examples is both tractable and relatively rich, and serves as a sample case for all applications that follow. Torus correlators for current algebras were of course considered before, see \cite{Mathur:1988jg,Dolan:2007eh}, and we focus on the $\tau\to0$ limit. We use dimensionless coordinate $z$ on the torus,
\begin{equation}
z \sim z+ 2\pi \sim z + 2\pi\tau.
\end{equation}
It is convenient to introduce a length scale $\ell$, so the OPE is written as
\begin{equation}
J_A(z_1)J_A(z_2) \sim \frac{k\frac{\psi^2}{2}\delta_{AB}}{\ell^2 (z_1-z_2)^2} + \frac{i f_{AB}{}^{C} J_C(z_2)}{\ell (z_1 - z_2)},
\end{equation}
where $\delta_{AB}$ is the invariant metric on $\mathfrak{g}$, and $\psi^2$ is the squared long root of a simple Lie algebra $\mathfrak{g}$. The case of abelian $\mathfrak{g}$ is slightly special because there is no $\psi^2$, nor canonical normalization for abelian currents, but it can be formally included in this analysis by picking \emph{some} normalization.

We now determine the two- and three-point functions following \cite{Mathur:1988jg}. By symmetry, $\langle J_A\rangle=0$ and the two-point function is proportional to $\delta_{AB}$, while the OPE implies it only has the second order pole, which is easily matched by the Weierstrass function:
\begin{equation}
\langle J_A(z_1) J_B(z_2)\rangle = \frac{k\psi^2}{2\ell^2} \delta_{AB}\left( \frac1{(2\pi)^2}\wp\left( \frac{z_1-z_2}{2\pi},\tau \right) + e(\tau) \right).
\end{equation}
The remainder term $e(\tau)$ is holomorphic and thus a constant, depending only on the complex structure of the torus. Because $\wp(z)$ has no constant term in its Laurent expansion, $e(\tau)$ is determined through the Sugawara construction,
\begin{equation}
\label{e_T}
\frac{k\psi^2}{2\ell^2}|G|e(\tau)= \sum_A\oint \frac{\dd w}{2\pi i w} \langle J_A(z+w) J_A(z)\rangle = \psi^2(k+h)\langle T_{\rm Sug}(z) \rangle .
\end{equation}
The stress tensor one point function is
\begin{equation}
\label{T_Z}
\langle T_{\rm Sug}\rangle = -\frac1{\ell^2}\frac{\dd \log Z}{\dd\log q} = -\frac1{2\pi i\ell^2} \frac{\dd \log Z}{\dd\tau} ,
\end{equation}
where $Z$ is the torus partition function. 

In terms of the genus-1 Szeg\"o kernels \cite{Mathur:1988jg}, 
\begin{equation}
\label{S_def}
S_i(z|\tau)=\frac{\theta_1'(0;\tau)\theta_i(z;\tau)}{\theta_i(0;\tau)\theta_1(z;\tau)},\quad i=2,3,4,
\end{equation}
which obey
\begin{equation}
\left[ S_i(z|\tau)\right]^2 = \wp(z;\tau)-e_i(\tau),\quad e_i(\tau) = -4\pi i \frac{\dd}{\dd\tau} \ln \frac{\theta_i(0;\tau)}{\eta(\tau)},
\end{equation}
where $\eta(\tau)$ is the standard Dedekind function, the two-point function can be written as 
\begin{equation}
\langle J_A(z_1) J_B(z_2)\rangle = \frac{k\psi^2}{2\ell^2} \delta_{AB} \sum_{i=2,3,4} W_i(\tau) \left[S_i\left( \frac{z_1-z_2}{2\pi};\tau \right) \right]^2,
\end{equation}
where
\begin{equation}
\label{Wi_s}
\sum_i W_i(\tau) = \frac1{(2\pi)^2},\quad \sum_i W_i(\tau) e_i(\tau) = -e(\tau).
\end{equation}
The three-point function has a simple expression in terms of these as well,\footnote{Equations \eqref{Wi_s} leave a one-parameter freedom in $W_i$'s, which does not affect the current two- and three-point functions \cite{Mathur:1988jg}.}
\begin{equation}
\langle J_A(z_1) J_B(z_2) J_C(z_3) \rangle = -\frac{i\psi^2k f_{ABC}}{4\pi\ell^3} \sum_i W_i S_i\left(\frac{z_{12}}{2\pi}\right)S_i\left(\frac{z_{23}}{2\pi}\right)S_i\left(\frac{z_{31}}{2\pi}\right).
\end{equation}

\paragraph{The $\tau\to 0$ limit.}
Let us now study the $\tau\to0$ limit of torus correlators. The $\tau\to 0$ asymptotic behavior of $\wp(z;\tau)$ follows from:
\begin{equation}
\int_0^\tau\dd w\, \wp(z+w,\tau) = -\frac1{\tau}G_2(-\frac1{\tau}) = -\frac{\pi^2}{3\tau} + O(e^{-\frac{2\pi i}{\tau}}),
\end{equation}
where $G_2$ is the first Eisenstein series, which implies that
\begin{equation}
\wp(z;\tau)=-\frac{\pi^2}{3\tau^2} + \text{exponentially small corrections}.
\end{equation}
Using expressions \eqref{S_def} and the $\tau\to0$ behavior of theta-functions, it is straightforward to obtain
\begin{align}
S_2(z;\tau) &= 0 + O(e^{-i\frac{\#}{\tau}})\,,\cr
S_3(z;\tau) &= 0 + O(e^{-i\frac{\#}{\tau}})\,,\cr
S_4(z;\tau) &= \frac{i\pi}{\tau}{\rm Sgn}\left({\rm Re}(z) \right) + O(e^{-i\frac{\#}{\tau}})\,.\cr
\end{align}
To fully describe the $\tau\to 0$ asymptotics of the two- and three-point functions, it remains to determine that of $e(\tau)$, which, unlike $S_i$'s, depends on the precise choice of the module we use to define torus correlators. Since we are studying the vacuum torus correlators, $e(\tau)$ is determined via \eqref{e_T},\eqref{T_Z}, with $Z$ the vacuum character.

The vacuum character equals the Schur index of the parent 4d SCFT, and at least for $c_{\rm 4d}> a_{\rm 4d}$, the $\tau\to 0$ behavior of the latter is given by \cite{DiPietro:2014bca,Ardehali:2015bla,Beem:2017ooy,Chang:2019uag}
\begin{equation}
\label{Cardy_Z}
\log Z \sim \frac{4\pi i (c_{\rm 4d}-a_{\rm 4d})}{\tau},
\end{equation}
implying the following behavior of $e(\tau)$:
\begin{equation}
e(\tau)\sim \frac{n}{\tau^2},\ \text{where } n=\frac{4(k+h^\vee)(c_{\rm 4d}-a_{\rm 4d})}{k \dim\mathfrak{g}}.
\end{equation}
Using this to solve \eqref{Wi_s}, we obtain
\begin{equation}
W_2+W_3 = \frac{1+6n}{6\pi^2},\quad W_4 = \frac{1-12n}{12\pi^2},
\end{equation}
resulting in the following $\tau\to0$ behavior of correlators:
\begin{align}
\langle J_A(a_1)J_B(z_2)\rangle &\sim -\frac{\psi^2}{2\ell^2\tau^2}\delta_{AB}\left( \frac{k}{12}-\frac{4(k+h^\vee)(c_{\rm 4d}-a_{\rm 4d})}{\dim\mathfrak{g}} \right),\cr\langle J_A(z_1)J_B(z_2)J_C(z_3)\rangle &\sim \frac{\psi^2 f_{ABC}}{4\ell^3\tau^3}\left( \frac{k}{12}-\frac{4(k+h^\vee)(c_{\rm 4d}-a_{\rm 4d})}{\dim\mathfrak{g}} \right){\rm Sgn}(Re(z_{12})){\rm Sgn}(Re(z_{23})){\rm Sgn}(Re(z_{13})).\cr
\end{align}
To have finite $\tau\to0$ limits, we ought to renormalize currents by $\tau$,
\begin{equation}
j_A \equiv -i\tau J_A.
\end{equation}
The correlators of $j_A$ we obtain are topological, depending only on the ordering of operators, as expected. The two-point function determines the metric, and the three-point function encodes the non-commutative associative star-product,\footnote{Note that we assume $f^{ACD}f^B{}_{CD}=h^\vee \psi^2\delta^{AB}$, therefore $\psi^2$ determines the normalization of generators.}
\begin{equation}
\label{star_AKM}
j_A \star j_B = :j_A j_B: + \frac{i\hbar}{2}f_{AB}{}^C j_C + \frac{\hbar^2}{4}\mu\psi^2 \delta_{AB}, \text{ where } \mu = \frac{k}{6}-\frac{8(k+h^\vee)(c_{\rm 4d}-a_{\rm 4d})}{\dim\mathfrak{g}}.
\end{equation}
We have introduced two new notations here. One is $\hbar=\ell^{-1}$, and the other is $:j_Aj_B:$, which is simply defined as the dimension-two operator appearing in $j_A\star j_B$ that is orthogonal to all lower-dimension operators. This $:j_A j_B:$ can be thought of as the normal ordering in 1d (related to normal ordering in 3d), and it differs from the VOA normal ordering $(j_A j_B)$ via mixing with the lower-dimension operators $j_A$ and the identity $1$. 
In the 3d CFT, $\mu$ is related to the flavor central charge $C_J$ of the global symmetry $\mf{g}$ by
\ie
\mu=-{C_J \over 32}.
\fe

Let us compare this to the formula obtained in \cite{Dedushenko:2019mzv}. There, the same star-product was derived from modularity, with the scalar parameter given by
\begin{eqnarray}
\mu = \frac{4(k+h^\vee)\widetilde{\Delta}_{\rm min}}{\dim \mathfrak{g}}.
\end{eqnarray}
The derivation we presented here is more explicit but uses the asymptotic behavior of the Schur index. The two answers agree if
\begin{equation}
\widetilde{\Delta}_{\rm min} = \frac{k\dim\mathfrak{g}}{24(k+h^\vee)} - 2(c_{\rm 4d}-a_{\rm 4d}) = \frac{c_{\rm 2d}}{24} - 2(c_{\rm 4d}-a_{\rm 4d}), 
\end{equation}
where $c_{\rm 2d} = c_{\rm Sug}$ is the 2d Sugawara central charge. This $\widetilde{\Delta}_{\rm min}$ agrees with $\widetilde{h}_{\rm min}$ from \cite{Beem:2017ooy}.

The algebra \eqref{star_AKM} describes quantization of the minimal nilpotent orbit of $\mathfrak{g}$, and is expected to apply to theories whose Higgs branch is the latter. Such theories provide the simplest non-trivial examples of our construction, and at the same time serve as a base for more involved applications to theories with the W-algebra chiral symmetry.

\section{Minimal nilpotent orbits}\label{sec:MinimalNilp}
In this section, we consider theories whose chiral algebra is the affine VOA of some simple $\mathfrak{g}$. Examples include the Deligne-Cvitanovi\'c (DC) exceptional series of rank-1 theories and the $(A_1, D_{2n+1})$ Argyres-Douglas theories (of which $(A_1, D_3)$ also belongs to the DC series), which share the common feature that the Higgs branch is described by the minimal nilpotent orbit of $\mf{g}$.

All the DC theories except the $(A_1, D_3)$ have a simplifying property that their UV and IR R-charges match. Namely, a charge $(R,r)$ of the $SU(2)_R\times U(1)_r$ representation in the UV coincides with the charge of representation of the enhanced IR R-symmetry $SU(2)_H\times SU(2)_C$
 \ie
 R_R=R_H,\quad r=R_C
 \fe
 where we adopt the convention that the $SU(2)$ spins $R_C,R_H,R_R \in \bZ/2$ and $r=\pm 1/2$ for the 4d supercharges.
 
  We refer to this case as ``no mixing''. On the contrary, for the $(A_1, D_{2n+1})$ theories, the $U(1)_r$ R-charges are fractional and cannot match the $SU(2)_C$ R-charges: the latter are given by mixing of the former with the Coulomb branch symmetries \cite{Buican:2015hsa}. We refer to this as the ``mixing'' case, which is expected to be generic in the space of 4d $\cN=2$ SCFTs.
\subsection{Without R-symmetry mixing}
Minimal 3d $\cN=4$ theories with $\mathfrak{g}$ flavor symmetry (acting on the Higgs branch given by the minimal nilpotent orbit of $\mathfrak{g}$) were recently considered in \cite{Chang:2019dzt}. In particular, they determine quantizations of the minimal nilpotent orbits, which describe the 1d protected sectors of these theories, from the bootstrap approach. The corresponding quantum algebra is a quotient of the universal enveloping algebra $\cU(\mathfrak{g})$ over the Joseph ideal, -- see also the discussion in \cite{ERS1}. For algebras different from $A_n$, this ideal is unique, and so is the quantization; for $A_n$, $n\geq2$, there is a one-parameter family of quantizations, only one of which is even; finally, for the spacial case of $\mathfrak{g}=A_1$, there is a one-parameter family of even quantizations (see  \cite{ASTASHKEVICH200286,Fronsdal2009,losev2015quantizations}). We focus on the value of the quadratic Casimir, which is proportional to $\mu$ as in the previous section. For $\mf{g}=A_1$, this $\mu$ parametrizes the family of quantization.\footnote{The corresponding star product is non-degenerate as long as $\mu$ stays away from the values corresponding to finite-dimensional representations of $A_1$, which in our normalization are $\mu=n(n+2)/6, n\in\Z_{\geq 0}$ \cite{ERS1}.} The value of $\mu$ that follows from the bootstrap approach is read off from the Table 6 in \cite{Chang:2019dzt} (it is twice their $\lambda_2$):
\begin{table}[!htb]
	\begin{center}
		\begin{tabular}{ |c|c| c| c|c|c| c|c|c|c|}
			\hline
			$\mathfrak{g}$ & $A_{n-1},\,n\ge3$ & $B_n$ & $C_n$ & $D_n$ & $E_6$  & $E_7$  & $E_8$  & $F_4$  & $G_2$  
			\\
			\hline
			$\mu $ & $-{  n\over 2(n+1)}$ &  $-{2n-3\over 2n}$   & $-{1\over 4}$ &  $-{2(n-2)\over 2n-1}$  & $-{12\over 13}$ & $-{24\over 19}$ & $-{60\over 31}$ & $-{3\over 4}$ & $-{4\over 9}$
			\\\hline
		\end{tabular}
	\end{center}
	\caption{$\mu$ from the 3d bootstrap.}
	\label{tab:dqm}
\end{table}

\subsubsection{DC theories excluding the $A_1$ case}
We expect all theories from the Table \ref{tab:dqm} to exist as 3d $\cN=4$ SCFTs. Indeed, the $ABCDE$ type 3d theories all have quiver gauge theory descriptions in the UV \cite{Douglas:1996sw}. While for the $G_2$ and $F_4$ the explicit constructions are missing, there are no known obstructions to their existence as well.

Their 4d lifts, however, do not always exist. Perturbative anomaly considerations restrict $\mathfrak{g}$ to the DC exceptional series (apart from the free hyper case). Global anomaly matching further rules out the $F_4$, and the fate of $G_2$ is not clear yet \cite{Shimizu:2017kzs}. Therefore we are left with the following list of 4d ``minimal'' theories with $\mathfrak{g}$ symmetry (we also do not consider the $A_0$ case which corresponds to the empty Higgs branch):  $A_1, A_2, D_4, E_6, E_7, E_8$. The $A_1$ theory exhibits mixing and thus will be considered later. Here we look at the remaining five cases.

The $A_2$ theory has as chiral algebra the affine VOA $V_{-3/2}(A_2)$, the $D_4$ theory has the chiral algebra $V_{-2}(D_4)$, and the $E_{6,7,8}$ Minahan-Nemeschansky (MN) theories have the chiral algebras $V_{-3}(E_6)$, $V_{-4}(E_7)$ and $V_{-6}(E_8)$, respectively. We summarize the relevant parameters and the resulting $\mu$, computed using \eqref{star_AKM}, in the following table:
\begin{table}[!htb]
	\renewcommand{\arraystretch}{1.7}
	\begin{center}
		\begin{tabular}{ |c|c| c| c|c|c| c|}
			\hline
			$\mathfrak{g}$ & $\dim\mathfrak{g}$ & $h^\vee$ & $k$ & $c_{\rm 4d}$ & $a_{\rm 4d}$  & $\mu$   
			\\
			\hline
			$A_2 $ & $8$ &  $3$   & $-3/2$ &  $\frac{2}3$  & $\frac{7}{12}$ & $-\frac{3}{8}$
			\\
			\hline
			$D_4 $ & $28$ &  $6$   & $-2$ &  $\frac{7}6$  & $\frac{23}{24}$ & $-\frac{4}{7}$
			\\
			\hline
			$E_6 $ & $78$ &  $12$   & $-3$ &  $\frac{13}6$  & $\frac{41}{24}$ & $-\frac{12}{13}$
			\\
			\hline
			$E_7$ & $133$ & $18$ & $-4$ & $\frac{19}{6}$ & $\frac{59}{24}$ & $-\frac{24}{19}$
			\\
			\hline
			$E_8$ & $248$ & $30$ & $-6$ & $\frac{31}6$ & $\frac{95}{24}$ & $-\frac{60}{31}$
			\\\hline
		\end{tabular}
	\end{center}
	\caption{$\mu$ for DC theories from the VOA.}
	\label{tab:E678}
\end{table}
\\
This perfectly matches the 3d results from Table \ref{tab:dqm}.\footnote{Note that the conformal and flavor central charges for the candidate $F_4$  and $G_2$ theories were determined in \cite{Beem:2013sza,Beem:2017ooy}, which after using \eqref{star_AKM} do give the expected answers for the 3d theories as in  Table \ref{tab:dqm}. Even though the 4d $F_4$ theory is ruled out in \cite{Shimizu:2017kzs} and the existence of the 4d $G_2$ theory is unclear, we see that at the level of the 2d chiral algebra, the $S^1$ reduction does give the expected TQM as predicted by the general argument in \cite{Dedushenko:2019mzv}.
} Below, we give a slightly more detailed treatment of the $D_4$ and $A_2$ cases. 

\subsubsection{$SU(2)$ SQCD, or more on the $D_4$ DC theory}
The $D_4$ DC theory has a simple Lagrangian description as the 4d $\cN=2$ $SU(2)$ gauge theory with $N_f=4$ fundamental hypermultiplets, whose chiral algebra is indeed $V_{-2}(\mathfrak{so}(8)) = V_{-2}(D_4)$ \cite{Beem:2013sza}. Dimensional reductions of this theory is the 3d gauge theory with the same gauge group $SU(2)$ and the same number of fundamental hypermultiplets. One can then apply the techniques of \cite{Dedushenko:2016jxl} to determine the value of $\mu$ from localization, which can be compared to $\mu=-{4\over 7}$ given above, testing both the 3d bootstrap and the VOA answers.

In this particular case, it is convenient to think of the 4d theory in terms of half-hypers, and the result of \cite{Dedushenko:2016jxl} --- the matrix model coupled to quantum mechanics that captures the 1d sector --- can be presented in the following form,
\begin{equation}
Z=\frac12\int\dd\sigma\, 4\sinh^2(\pi\sigma)\int \cD X e^{-S_{\rm 1d}}.
\end{equation}
Here the 1d action is written in terms of fields $X_{i\alpha}(\varphi)$, $i=1\dots 8$, $\alpha=1,2$ on the circle as
\begin{equation}
S_{\rm 1d}=-\frac1{2\hbar}\int_{-\pi}^\pi \dd\varphi\, \varepsilon^{\alpha\beta}X_{i\beta}\left(\partial_\varphi X_i +\sigma \frac{\tau^3}{2} X_i \right)_\alpha,
\end{equation}
where $\tau^i$ denotes Pauli matrix acting on the gauge indices $\alpha,\beta,\dots$, $\hbar$ is related to the radius of the sphere via $\hbar=1/(8\pi\ell)$, and we use the convention $\varepsilon^{12}=\varepsilon_{21}=+1$. The correlators of gauge-invariant operators are topological (depend only on the order of operator insertions), so it is enough to know the Green's function (for a given value of $\sigma$) at $\epsilon\to0$ only,
\begin{equation}
\label{GreenD4}
\langle X_{i\alpha}(\epsilon)X_{j\beta}(0) \rangle_\sigma \sim -\frac{\hbar}{2} \delta_{ij}\left[ {\rm Sgn}(\epsilon)\varepsilon_{\beta\alpha} + \tanh\frac{\pi\sigma}{2} (\tau_1)_{\alpha\beta} \right].
\end{equation}
The $\mathfrak{so}(8)$ generators are defined by
\begin{equation}
J_{ij}=i\varepsilon^{\alpha\beta} X_{[i\beta}X_{j]\alpha} = i\varepsilon^{\alpha\beta}\lim_{\epsilon\to+0} \frac{X_{i\beta}(\epsilon)X_{j\alpha}(0)-X_{j\beta}(\epsilon)X_{i\alpha}(0)}{2}.
\end{equation}
Using \eqref{GreenD4} and Wick contractions, it is straightforward to find the following correlators,
\begin{align}
\langle J_{ij}(\epsilon)J_{km}(0)\rangle &= - \left( \frac{\hbar}{2}\right)^2 (\delta_{ik}\delta_{jm} - \delta_{im}\delta_{jk})\left\langle \frac2{\cosh^2 \frac{\pi\sigma}{2}}\right\rangle,\cr
\langle J_{ij}(\epsilon+\mu)J_{km}(\epsilon)J_{pq}(0)\rangle_{\epsilon,\mu >0} &= \left( \frac{i\hbar}{2} \right)^3 \Big( \delta_{ik}(\delta_{jp}\delta_{mq}-\delta_{jq}\delta_{mp}) -\delta_{im}(\delta_{jp}\delta_{kq}-\delta_{jq}\delta_{kp})\cr 
&- \delta_{jk}(\delta_{ip}\delta_{mq}-\delta_{iq}\delta_{mp}) + \delta_{jm}(\delta_{ip}\delta_{kq}-\delta_{iq}\delta_{kp}) \Big)\cdot \left\langle \frac2{\cosh^2 \frac{\pi\sigma}{2}}\right\rangle,
\end{align}
where
\begin{equation}
\left\langle \frac2{\cosh^2 \frac{\pi\sigma}{2}}\right\rangle = \frac{\int\dd\sigma \frac{\sinh^2\pi\sigma}{(\cosh^2(\pi\sigma/2))^4} \cdot \frac{2}{\cosh^2(\pi\sigma/2)}}{\int\dd\sigma \frac{\sinh^2\pi\sigma}{(\cosh^2(\pi\sigma/2))^4}}=\frac{8}{7}.
\end{equation}
These results imply the following star-product of currents,
\begin{equation}
J_{ij}\star J_{km} = :J_{ij}J_{km}: + \frac{i\hbar}{2} (\delta_{ik}J_{jm}-\delta_{im}J_{jk}+\delta_{jm}J_{ik}-\delta_{jk}J_{im}) + \frac{\hbar^2}{4} \times \left(-\frac87\right)(\delta_{ik}\delta_{jm}-\delta_{im}\delta_{jk}),
\end{equation}
where $:J_{ij}J_{km}:$ by definition is a dimension-2 operator that appears in this product and is orthogonal to all lower-dimension operators. This product is of the form \eqref{star_AKM}: the $O(\hbar)$ term contains the $\mathfrak{so}(8)$ structure constants in the $\psi^2=2$ normalization, and the $O(\hbar^2)$ term implies
\begin{equation}
\mu=-\frac47,
\end{equation}
matching the value in the Table \ref{tab:E678}.

\subsubsection{$(A_1, D_4)$ AD, or more on the $A_2$ DC theory}
The $A_2$ DC theory coincides with the $(A_1, D_4)$ Argyres-Douglas (AD) theory. Though non-Lagrangian in 4d, its 3d reduction is known to admit a Lagrangian description \cite{} as an SQED${}_3$, whose 3d mirror is also Lagrangian and given by a $U(1)\times U(1)$ gauge theory with hypers of charges $(1,0)$, $(1,1)$ and $(0,1)$ (this is also equivalent to a 3-node necklace quiver).

We may use either of the available 3d descriptions to find the protected algebra. In particular, the SQED${}_N$ for general $N$ was studied in details in \cite{Dedushenko:2016jxl}, and the $N=3$ specialization is described by the algebra
\begin{equation}
\cJ_I{}^J\star \cJ_K{}^L=\cJ_{IK}{}^{JL} + \frac{i\hbar}{2} (i\delta_K^J\cJ_I{}^L - \delta_I^L\cJ_K{}^J) + \frac{\hbar^2}{4} \times \left(-\frac{3}{4} \right)\left(\delta_I^L\delta_K^J - \frac13 \delta_I^J\delta_K^L \right),
\end{equation}
where $\cJ_I{}^J$ should be thought of as traceless complex $3\times 3$ matrices (\emph{i.e.}, the generators of $\mathfrak{sl}_3$). Again being careful about the normalization, we conclude that $\psi^2=2$ and
\begin{equation}
\mu=-\frac{3}{8},
\end{equation}
in agreement with the corresponding entry in Table \ref{tab:E678}. One can also obtain the same answer from the Coulomb branch computation in the mirror dual $U(1)\times U(1)$ gauge theory using the technique developed in \cite{Dedushenko:2018icp}, see Appendix \ref{app:ShiftOp}.

\subsection{With R-symmetry mixing}
In general, when a 4d $\cN=2$ SCFT is reduced on $S^1$, the Higgs branch is unrenormalized, whereas the Coulomb branch gets enhanced by extra fiber coordinates \cite{Seiberg:1996nz} from 4d BPS line operators wrapping the $S^1$ factor. In the 3d limit, this is often accompanied by the appearance of accidental $U(1)$ symmetries associated to the topological currents on the 3d Coulomb branch, which we refer to as Coulomb branch symmetries. The Coulomb branch symmetries can enter mixing relations that identify the 4d $U(1)_r$ current as the combination of 3d currents that involve the Cartan of $SU(2)_C$ and the Coulomb branch symmetries.
As explained in \cite{Buican:2015hsa}, this happens when the 4d theory contains Coulomb branch chiral primaries with $r\notin \frac12 \bZ$.

The supersymmetric $S^1\times S^3$ background couples to the 4d $U(1)_r$ symmetry. Therefore, in the 3d limit, we obtain the $S^3$ background coupled to $U(1)_r$, which is not the right R-symmetry due to mixing. This background differs from the supersymmetric $S^3$ coupled to the conformal R-symmetry only by the presence of imaginary masses for symmetries that participate in mixing \cite{Festuccia:2011ws}. Such imaginary masses for the Coulomb branch (or topological) symmetries are equivalent to imaginary FI parameters for abelian gauge symmetries. They can deform the Higgs branch TQM in 3d, and must be taken into account in relation to the VOA. If the generator $r$ of $U(1)_r$ and the Cartan element $R_C$ of $SU(2)_C$ are related through mixing with the topological charges $\cT^1,\dots, \cT^m$,
\begin{equation}
\label{mixing_ralation}
r = R_C+ \sum_{a=1}^m c_a \cT^a,
\end{equation}
then the FI parameters for the corresponding $U(1)_a$ gauge factors in 3d are simply given by \cite{Buican:2015hsa}
\begin{equation}
\zeta_a = \frac{i}{\ell}c_a.
\end{equation}

Each of the $(A_1,D_{2n+1})$  Argyres-Douglas theories has a different Coulomb branch of complex dimension $n$, but all share the same Higgs branch $\bC^2/\bZ_2$ (the minimal nilpotent orbit for $\mf{g}=A_1$). The corresponding chiral algebras are $V_{-{4n\over 2n+1}}(\mf{sl}_2)$.  These theories have fractional Coulomb branch spectrum given by $2i\over 2n+1$ with $i=n+1,n+2,\dots, 2n$, which is why they exhibit mixing.

\subsubsection{$(A_1, D_3)$ AD, or the $A_1$ DC theory}

In the special case of $n=1$, the 4d SCFT belongs to the DC series. The corresponding chiral algebra is $V_{-{4\over3}}(\mf{sl}_2)$. This reduces to a 3d TQM via \eqref{star_AKM} which is a universal enveloping algebra $\cU(A_1)$  (with central quotient) with 
\ie
\label{AD3chirAnswer}
\mu=-{8\over 27}.
\fe
On the other hand, the 3d SCFT from $S^1$ reduction is described by $\cN=4$ SQED with 2 unit-charge hypermultiplets, usually called $T[SU(2)]$. The TQM of this theory was solved in \cite{Dedushenko:2016jxl}, and  
\ie
\label{AD3TQMAnswer}
\mu^{3d}=-{1\over 3}
\fe
As we have mentioned, the discrepancy between \eqref{AD3chirAnswer} and \eqref{AD3TQMAnswer} comes from the nontrivial R-symmetry mixing,  which in this case is given in  \cite{Buican:2015hsa} by
\ie
\label{ad3Mixing}
r = R_C+{1\over 3}\cT,
\fe
where $\cT$ denotes the accidental $U(1)$ symmetry (enhanced to $SU(2)$) on the 3d Coulomb branch. The corresponding FI deformation is given by $\frac{i}{\ell}$ times the coefficient in the mixing relation \eqref{ad3Mixing},
\ie
\zeta={i\over 3\ell}.
\label{D3fi}
\fe
Recall the 3d TQM from the 3d SQED$_2$ with a general FI parameter $\zeta$ from \cite{Dedushenko:2016jxl},
\ie
\cJ_I{}^J\star \cJ_K{}^L=&\cJ_{IK}{}^{JL}
- {1\over 2\ell}(\D^J_K \cJ_I{}^L-\cJ_K{}^J \D_I^L)
-{\zeta^2 \ell^2+1\over 6 \ell^2} (\D^L_I \D_K^J-{1\over 2}\D^J_I \D^L_K),
\fe
where $\frac{1}{\ell}=\hbar$ puts algebra into the canonical form. With the FI parameter value in \eqref{D3fi},
\ie
\mu=-{1+\zeta^2\ell^2\over 3}=-{8\over 27},
\fe
in agreement with the answer \eqref{AD3chirAnswer} from reducing the chiral algebra.

\subsubsection{General $(A_1, D_{2n+1})$ AD and free twisted hypers}
\label{sec:A1Dodd}
For general $n$, 
the $(A_1, D_{2n+1})$ SCFT  has conformal central charges
\ie
a_{4d}={n(8n+3)\over 8(2n+1)},\quad c_{4d}={3n\over 6}.
\fe
In particular,
\ie
24(c_{4d}-a_{4d})={3n\over 2n+1},
\fe
which signals mixed branches. Applying \eqref{star_AKM}, the chiral algebra  $V_{-{4n\over 2n+1}}(\mf{sl}_2)$ reduced on $S^1$ also produces a TQM given by the central quotient of $\cU(\mathfrak{sl}_2)$, but with a different value of $\mu$, 
\ie
\label{A1DoddPrediction}
\mu=-\frac{4 n (n+1)}{3 (2 n+1)^2}.
\fe

The 3d theory from  the $S^1$ reduction (or its mirror dual, the so-called 3d mirror) is not known for the $(A_1, D_{2n+1})$  theories. A natural proposal that we make and test below is that
\begin{align}
\label{ReductionProposal}
(A_1,D_{2n+1}) \to \underbrace{h\otimes \dots \otimes h}_{n-1\text{ times}}~\otimes {\rm SQED}_2,\cr
\text{where } h= \left\{\text{free twisted hyper}\right\},
\end{align}
and ${\rm SQED}_2$ appears with an FI deformation that generalizes \eqref{D3fi},
\ie
\zeta= {i\over (2n+1)\ell}.
\label{A1DoddFI}
\fe

 In other words, we propose that the $(A_1, D_{2n+1})$ theory reduces to $n$ decoupled sectors, one of which is the interacting theory SQED$_2\equiv T[SU(2)]$, while the others are free twisted hypers. We will give more evidence below.
 Only the SQED$_2$ contributes to the Higgs sector, and ensures that the FI deformed TQM is given by the central quotient of $\cU(\mathfrak{sl}_2)$ with
 \ie
 \mu=-{1+\zeta^2\ell^2\over 3}=-\frac{4 n (n+1)}{3 (2 n+1)^2},
 \fe 
 matching the prediction \eqref{A1DoddPrediction} from the chiral algebra. In Section~\ref{sec:checkFTH}  we will elaborate more on \eqref{ReductionProposal}, provide further checks, and comment on a related issue of reducing the $(A_1, A_{2n})$ theories.

 \section{From interacting 4d SCFT to free fields  in 3d}\label{sec:checkFTH}
 In this section, we provide further evidence for the proposal \eqref{ReductionProposal} on the $S^1$ reduction of the $(A_1, D_{2n+1})$ theory. Furthermore, we explain why $h$ in \eqref{ReductionProposal} is precisely the twisted hyper, not its discrete gauging. It proves helpful to study this problem in conjunction with the reduction of the $(A_1, A_{2n-2})$ theory on the circle, as their class S constructions are related by Higgsing, and the $S^1$ reductions have very similar features, despite $(A_1, A_{2n-2})$ theories having no Higgs branch. Furthermore, while the 3d reduction and 3d mirrors of the $(A_1, A_{2n-1})$ and $(A_1, D_{2n})$ theories have received attention in the literature \cite{Nanopoulos:2010bv,Xie:2012hs,Buican:2015hsa}, we cannot say the same about $(A_1, A_{2n-2})$ and $(A_1, D_{2n+1})$.\footnote{See, however, \cite{Benvenuti:2018bav} on $(A_1, A_{2n-2})$.} This makes the question interesting on its own, even outside the present context.
 
 In Class S construction \cite{Gaiotto:2009we}, the $(A_1,D_{2n+1})$ theory is realized by twisted compactification of the $A_1$ $(2,0)$ theory on a sphere with one regular puncture and one irregular puncture  \cite{Gaiotto:2009hg,Xie:2012hs,Wang:2015mra, Wang:2018gvb}.With $z$ a complex coordinate on the sphere, the irregular puncture is described by a singularity of the Higgs field at $z=0$  of the following form,
 \ie
 \Phi(z) = {T\over z^{n+{3\over 2}}}+{\rm regular},
 \label{A1irregular}
 \fe
 where $T$ is a regular semisimple element of $\mf{su}(2)$. The Seiberg-Witten (SW) curve is given by
 \ie
 x^2+{z^{2n+1}+m^2+\sum_{a=1}^{n} u_a z^{n+1-a}+\sum_{a=1}^n v_a z^{n+a-1}\over z^2}=0,
 \fe
 with the SW differential $\lambda=xdz$. Here $m$ denotes the mass parameter for the $\mf{su}(2)$ flavor symmetry, and $u_a$  is (the vev of) the Coulomb branch chiral primary of dimension $\Delta_a=1+{2a-1\over 2n+1}$, while $v_a$ is the corresponding chiral coupling with dimension ${2(n-a+1)\over 2n+1}$.
 
 A closely related Class S setup that only involves the above irregular singularity \eqref{A1irregular} on the sphere describes the $(A_1,A_{2n-2})$ theory. The SW curve becomes
 \ie
 x^2+{z^{2n-1}+\sum_{a=1}^{n-1} u_a z^{n-1-a}+\sum_{a=1}^{n-1} v_a z^{n-3+a}}=0,
 \fe
 with the same SW differential $\lambda=xdz$. We have labeled the parameters in the SW curve  purposefully so that it is obvious that the Coulomb branch spectrum of the $(A_1,A_{2n-2})$ theory is identical to that of $(A_1,D_{2n+1})$, apart from the additional operator vev $u_n$ with $\Delta_n=2-\frac{2}{2n+1}$, and coupling $v_n$ that $(A_1, D_{2n+1})$ has. The latter pair comes from Hitchin moduli of the regular puncture in the Class S setup.
 
 At the physical level, the two theories are related by Higgsing, namely by giving a vev to the moment map operator associated to the $\mf{su}(2)$ flavor symmetry of $(A_1,D_{2n+1})$. In the class S setup, it is known that Higgsing corresponds to closing (reducing) regular punctures \cite{Chacaltana:2012zy}. When we reduce these theories on $S^1$, since moment map operators (or any Higgs branch chiral primary) are unambiguously identified between 4d and 3d, we expect the resulting 3d theories to be related by the same Higgsing as well. Together with our proposal \eqref{ReductionProposal} for the $S^1$ reduction of $(A_1,D_{2n+1})$, we are lead to the prediction that 
 \begin{align}
 \label{ReductionProposalA1Ae}
 (A_1,A_{2n-2}) \to 
 \underbrace{h\otimes \dots \otimes h}_{n-1\text{ times}} ,
 \end{align}
 where $h$ is the same twisted hypermultiplet as in \eqref{ReductionProposal}. Moreover, as we shall show below, the relevant R-symmetry mixing is captured by the masses $\zeta_a$ for the $\otimes_{a=1}^{n-1} U(1)_a$ Coulomb branch (topological) symmetries, under which the free twisted hypers are charged. We always refer to masses for the Coulomb branch symmetries as FI parameters.\footnote{A twisted hyper is dual to SQED$_1$, in which case the mass literally corresponds to the FI term across duality.} They are given by
 \ie
 \zeta_a= {i\over \ell}{2a-1\over 2n+1},\quad 1\leq a \leq n-1,
  \label{A1AevenFI}
 \fe 
which holds for both the reduction of $(A_1,A_{2n-2})$ and  $(A_1,D_{2n+1})$ theories. Recall that the reduction of $(A_1, D_{2n+1})$ theory has one more topological symmetry with the FI term given in \eqref{A1DoddFI}.

\subsection{The reduction of $(A_1, A_{2n-2})$. }
The $(A_1, A_{2n-2})$ reduces to a 3d $\cN=4$ SCFT that has no Higgs branch, while the Coulomb branch acquires quaternionic dimension $n-1$. Correspondingly, the 3d mirror has no Coulomb branch and an $(n-1)$-dimensional Higgs branch. An obvious way to realize this scenario for the 3d mirror is by taking $n-1$ free hypermultiplets, or equivalently, $n-1$ free twisted hypermultiplets to describe the direct 3d reduction, which is precisely what happens, as we argue below (thus confirming the proposal of \cite{Benvenuti:2018bav}). We could also contemplate the possibility of discrete gaugings in this free system, but as we will see from the Coulomb branch index, this does not happen.

An $a$-th twisted hyper contributes two operators to the Coulomb branch chiral ring in 3d, which we denote as $q_a$ and $\widetilde{q}_a$. They both have dimension $\Delta=\frac12$, $SU(2)_C$ R-charge $R_C=\frac12$, and they form a doublet with respect to yet anther $SU(2)$. The topological symmetry is the maximal torus of the latter $SU(2)$, so the corresponding charges are $\pm\frac12$. We claim that these operators are emergent in the IR. The operator that directly flows from the corresponding  4d Coulomb branch chiral primary $u_a$ is, in fact, a composite operator $q_a q_a$ in 3d. This operator has $\Delta=R_C=1$, and the topological charge $\cT^a = 1$. Recall that $u_a$ has the 4d dimension and r-charge $\Delta_a=r_a=1+\frac{2a-1}{2n+1}$. This shows that the predicted mixing relation,
\begin{equation}
1+\frac{2a-1}{2n+1} = r_a = R_C + c_a \cT^a,
\end{equation}
indeed holds if the mixing coefficient is
\begin{equation}
c_a = \frac{2a-1}{2n+1},
\end{equation}
which is consistent with the imaginary FI parameter value stated in \eqref{A1AevenFI}.
 
We now proceed to provide more evidence for these claims. The  FI couplings \eqref{A1AevenFI} can be verified by studying directly the $S^1$ reduction of the 4d Schur index and comparing to (FI-deformed) $S^3$ partition of the proposed 3d SCFT.
 The Schur index of the $(A_1,A_{2n-2})$ theory is given by
 \ie
 \cI_{(A_1,A_{2n-2})}={\rm PE}\bigg[
 {q^2-q^{2n}
 	\over 
 	(1-q)(1-q^{2n+1})	
 }
 \bigg].
 \fe
 We would like to take the $\tau \to 0$ limit. A useful trick is to rewrite the plethystic exponential (PE) in terms of an ordinary exponential of a sum of Lambert series,
 \ie
 \cL_q(s,x)\equiv \sum_{k=1}^\infty {k^s q^{kx}\over 1-q^k },
 \fe
 which have simple behavior as $\tau \to 0$. The particular limit formula we need here is 
 \ie
 \cL_q(-1,x)=
 -{\zeta(2)  \over \log q}
 +\log {\Gamma(x)\over \sqrt{2\pi}}+ 
 \log \log {1\over q} \left(x-{1\over 2}\right)
 +\cO(\log q),
 \label{Lambertlim1}
 \fe
 which describes the $q\to 1$ behavior. By writing
 \ie
 \cI_{(A_1,A_{2n-2})}=&{\rm PE} \left[
 {q^2+q^3 +\dots+ q^{2n-1} \over 1-q^{2n+1}}
 \right]
 =
 &
 \exp \left[
 \sum_{j=2}^{2n-1}\cL_{q^{2n+1}} \left(-1,{j\over 2n+1}\right)
 \right]
 \fe
 and using \eqref{Lambertlim1}, we obtain
 \ie
 \log\cI_{(A_1,A_{2n-2})}=&
 {{  (n-1) \over 12(2n+1)} {2\pi i\over \tau} }
 +\sum_{j=2}^{2n-1} \log {\Gamma(j/(2n+1))\over \sqrt{2\pi}} +\cO(\log q).
 \fe
 The first term above diverges as $\tau \to 0$ and captures the Cardy limit of the index,   whose coefficient is determined in terms of the 4d conformal central charges as 
 \ie
 \lim_{\tau \to 0} (-i \tau)\log \cI_{\rm Schur} ={4\pi (c_{4d}-a_{4d}) },
 \label{4dcardyF}
 \fe according to the general arguments of \cite{DiPietro:2014bca,Ardehali:2015bla,Beem:2017ooy,Chang:2019uag}. This is clearly consistent with the central charges of the ${(A_1,A_{2n-2})}$ theory \cite{Xie:2013jc},
 \ie
 a^{(A_1,A_{2n-2})}_{4d}=\frac{(n-1) (24 n-5)}{24(2n+1)},\quad c^{(A_1,A_{2n-2})}_{4d}=\frac{(n-1) (6 n-1)}{6(2n+1)}.
 \fe

 We are interested in the finite piece, which determines the partition function of the 3d SCFT in the IR, and which we compare to the partition function of twisted hypers,
 \ie
 Z_{\rm 3d}
 =\prod_{j=2}^{2n-1} {\Gamma\left(j/ (2n+1)\right)\over \sqrt{2\pi}}.
 \fe
 This indeed naturally factorizes into a product of $n-1$ twisted hypermultiplet partition functions with the FI deformations (that give masses to the twisted hypers) as 
 \ie
 Z_{\rm 3d}=\prod_{a=1}^{n-1} Z_{\eta}(\zeta_a),
 \fe
 with
 \ie
 Z_{\eta}(\zeta)\equiv {\Gamma(1/2 + i \ell\zeta/2)\Gamma(1/2 - i \ell\zeta/2)\over 2\pi}=\frac1{2\cosh \pi\frac{\ell\zeta}{2}},
 \fe
 where $\frac12$ in $\frac{\ell\zeta}{2}$ originates from the topological charges of elementary fields, and the FI parameter is
 \ie
 \zeta_a= {i\over \ell}{2a-1\over 2n+1},
 \fe
 which confirms \eqref{A1AevenFI} for the $(A_1, A_{2n-2})$ theories.
 
\begin{figure}[t!]
	\centering
	\begin{subfigure}[b]{0.48\textwidth}
		\centering
		\includegraphics[width=\textwidth]{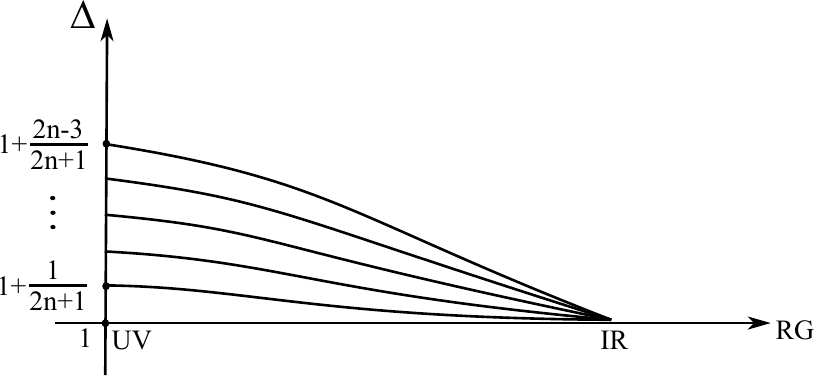}
		\caption{$(A_1, A_{2n-2})$: dimensions $\Delta$ flow to $1$.}
		\label{fig:Aeven}
	\end{subfigure}
	\hfill
	\begin{subfigure}[b]{0.48\textwidth}
		\centering
		\includegraphics[width=\textwidth]{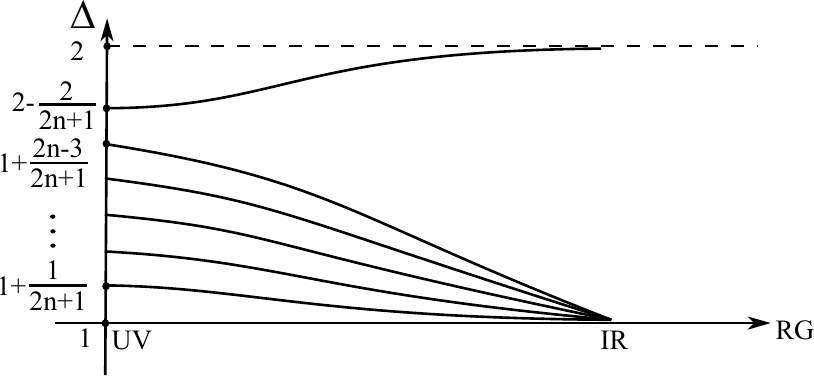}
		\caption{$(A_1, D_{2n+1})$: one operator flows to $\Delta=2$.}
		\label{fig:Dodd}
	\end{subfigure}
	\caption{Schematic RG flow of Coulomb branch chiral operators in $(A_1, A_{2n-2})$ and $(A_1, D_{2n+1})$ theories. The plot for $(A_1, D_{2n+1})$ contains $(A_1, A_{2n-2})$ as a subplot, and has one more special operator corresponding to the $T[SU(2)]$ in 3d.}
	\label{fig:Flow}
\end{figure} 
 
We see that our computations agree with the hypothesis that $(A_1, A_{2n-2})$ flows to the collection of $n-1$ free twisted hypers in 3d, and furthermore, that the 4d Coulomb branch operators $u_a$ flow to the dimension-1 composites $q_a q_a$, which is illustrated in Figure \ref{fig:Aeven}. However, we have not identified the elementary fields $q_a$ and $\widetilde{q}_a$, therefore have not excluded the possibility that the 3d theory actually consists of the $\Z_2$-gauged twisted hypers. Such a 3d theory would have the same $S^3$ partition function, would also contain operators like $q_a q_a$, but not $q_a$ or $\widetilde{q}_a$ individually.
 
 To distinguish free twisted hypers from their $Z_2$-gauged version, it is useful to consider quantities other than the Schur index, and backgrounds other than the round sphere. In particular, the lens spaces $L(p,1)=S^3/\Z_p$ give a useful family of backgrounds. For even $p$, the $\Z_2$ gauge theory can have a nontrivial holonomy around the 1-cycle of $L(p,1)$, thus introducing twisted sectors for the hypers. Therefore, partition functions on such backgrounds would differ for ordinary and $\Z_2$-gauged hypers by a factor of $2$. We could then use known results on the lens space indices (see, \emph{e.g.}, \cite{Benini:2011nc,Alday:2013rs,Razamat:2013jxa,Razamat:2013opa,Fluder:2017oxm}) to study the high-temperature limits thereof.
 
 Instead, we will do something even simpler, namely look at the 4d Coulomb branch index on lens spaces. On a round sphere, such an index simply counts Coulomb branch chiral operators $u_a$. Its lens space generalizations are known and have been computed in \cite{Fredrickson:2017yka} (see also \cite{Dedushenko:2018bpp} for generalizations). Furthermore, it is known that in the $p\to\infty$ limit, the lens space reduces to $S^2$ (the Hopf fiber shrinks to a point), and the $S^1 \times L(p,1)$ index reduces to the $S^1\times S^2$, \emph{i.e.} the usual 3d index \cite{Benini:2011nc}. In particular, if we start with the Coulomb branch index in 4d, we expect to obtain its analog in 3d, that is simply a generating function of the Coulomb branch chiral spectrum in the 3d $\cN=4$ theory we flow to. Here we implicitly assume that shrinking the thermal circle and shrinking the Hopf circle lead to the same 3d theory, which seems reasonable in local QFT.
 
 We also have to be careful and remember that in 4d, the Coulomb branch index counts states according to their conformal $r$-charge, and can be written as
 \begin{equation}
 \cI(t) = \trace_{C} (-1)^F t^{r},
 \end{equation}
 where $\trace_C$ denotes trace over the subspace of Coulomb branch states.\footnote{One can find other expressions in the literature, including the ones with $t^{r-R}$ or $t^{r+R}$. Since we view this simply as a counting function here, and since all the Coulomb branch operators have $R=0$, $t^r$ is enough for our purposes.} As we take the $p\to\infty$ limit, we will have to account for the $r$-symmetry mixing to get the correct answer in 3d. Thus we will not obtain the Coulomb index in 3d on the nose, but rather its close cousin,
 \begin{equation}
 \lim_{p\to \infty} \cI(t) = \trace_{C, 3d} (-1)^F t^{R_C + \sum_a c_a \cT^a},
 \end{equation}
where the trace goes over the space of 3d Coulomb branch operators, but the counting is affected by our familiar mixing. For the $(A_1, A_{2n-2})$ theories, the lens space index was obtained in \cite{Fredrickson:2017yka},
 \begin{equation}
 \cI_{n,p}(t) = \sum_{k=0}^{n-1} \frac{t^{\frac{k(k+1)}{2(2n+1)}p}}{\prod_{a=1}^k \left(1-t^{\frac{2(n+a)}{2n+1}}\right)\left(1-t^{-\frac{2a-1}{2n+1}}\right)\prod_{a=k+1}^{n-1}\left(1-t^{\frac{2a+1}{2n+1}}\right)\left(1-t^{\frac{2(n-a)}{2n+1}}\right)},
 \end{equation}
 where different terms in the sum correspond to fixed points of the Hitchin action on the wild Hitchin moduli space that arises from the class S construction of these theories. Taking the $p\to\infty$ limit, assuming that $|t|<1$, kills all the terms with $k>0$, and the only remaining term has $k=0$,
 \begin{equation}
 \label{CoulIndLimit}
 \lim_{p\to\infty} \cI_{n,p}(t) = \frac{1}{\prod_{a=1}^{n-1}\left(1-t^{\frac{2a+1}{2n+1}}\right)\left(1-t^{\frac{2(n-a)}{2n+1}}\right)}=\prod_{j=2}^{2n-1}\frac1{1-t^{\frac{j}{2n+1}}}.
 \end{equation}
 A simple computation gives the values of $R_C + \sum_a c_a\cT^a$ for the elementary hypermultiplet chirals,
 \begin{align}
 \text{For } q_a:\quad &R_C + c_a\cT^a = \frac12 (1+c_a) = \frac{n+a}{2n+1},\quad a=1\dots n-1,\cr
 \text{For } \widetilde{q}_a:\quad &R_C + c_a\cT^a = \frac12 (1-c_a) = \frac{n-a+1}{2n+1},\quad a=1\dots n-1,
 \end{align}
which are just $\frac{j}{2n+1}$ with $j$ ranging from $2$ to $2n-1$, in perfect agreement with \eqref{CoulIndLimit}. This computation gives us confidence that the operators $q_a$, $\widetilde{q}_a$ do belong to the 3d Coulomb branch spectrum, implying that we indeed obtain ordinary, not $\Z_2$-gauged, free twisted hypers. Notice that $q_a$ and $\widetilde{q}_a$ emerge in the 3d limit, perhaps from line operators wrapping the shrinking 1-cycle.

 \subsection{On 4d $\cN=4$ SYM with boundary conditions.}
\begin{figure}[t!]
	\centering
	\begin{subfigure}[b]{0.46\textwidth}
		\centering
		\includegraphics[width=\textwidth]{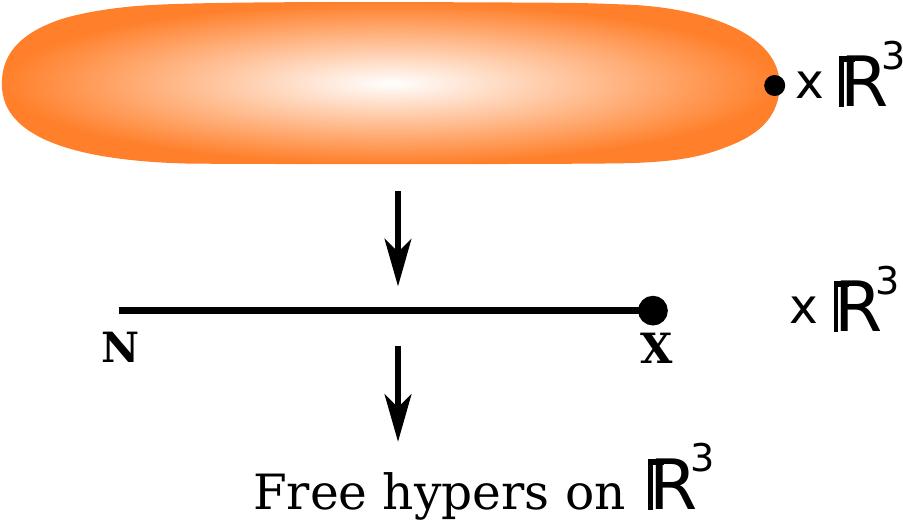}
		\caption{$(A_1, A_{2n-2})$ setup.}
		\label{fig:Aeven5d}
	\end{subfigure}
	\hfill
	\begin{subfigure}[b]{0.46\textwidth}
		\centering
		\includegraphics[width=\textwidth]{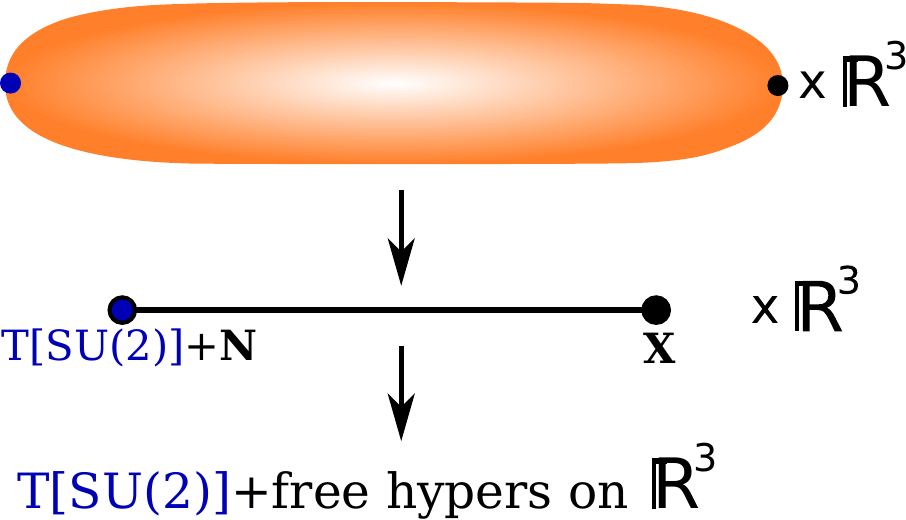}
		\caption{$(A_1, D_{2n+1})$ setup.}
		\label{fig:Dodd5d}
	\end{subfigure}
	\caption{Constructions of 3d mirrors for (a) $(A_1, A_{2n-2})$ and (b) $(A_1, D_{2n+1})$ from the 5d MSYM compactified on a cigar with one or two punctures, and the corresponding 4d MSYM on the interval.}
	\label{fig:MSYM-bc}
\end{figure}

Recall that class S theories are described by holomorphic compactifications of the 6d theory on a Riemann surface with punctures \cite{Gaiotto:2009hg,Nanopoulos:2009uw,Bonelli:2011aa,Xie:2012hs,Xie:2013jc,Wang:2015mra,Wang:2018gvb}. Since we further reduce to 3d on a circle, we can consider reversing the order of compactifications: this will first produce the 5d maximal super Yang-Mills (5d MSYM), which we then compactify on a Riemann surface. Due to peculiar properties of the 6d theory, this will actually result in the 3d mirror description of the same theory. Degenerating the Riemann surface, one can equivalently look at this as a 4d MSYM compactified on a certain graph, as was done in \cite{Benini:2010uu}, which builds up on the study of supersymmetric boundary conditions in 4d MSYM \cite{Gaiotto:2008sa,Gaiotto:2008ak}. Since for Argyres-Douglas theories the Riemann surface is just a sphere with one or two punctures, the corresponding graph becomes very simple, -- it is just an interval. Punctures determine what boundary conditions one has to impose at the endpoints of the interval. Therefore, one considers 4d MSYM reduced on an interval with specific boundary conditions, which flows to the 3d $\cN=4$ SCFT in the IR, see Figure \ref{fig:MSYM-bc}.
 
For the $(A_1, A_{2n-2})$ case that we considered so far, one starts with the $A_1$ theory in 6d, and there is only one irregular (or wild) puncture on the sphere. Therefore, we have the $SU(2)$ MSYM in 5d, and the reduction to 4d involves degenerating the cigar as depicted on Figure \ref{fig:Aeven5d}. One end of the cigar is completely empty, which is known to lead to the Neumann boundary condition (call it \textbf{N}) in the 4d degeneration limit \cite{Chacaltana:2012zy,Gaiotto:2008ak}. Another end involves the wild puncture, and the corresponding 4d boundary condition, --- denote it \textbf{X}, --- is not known. Nevertheless, our previous conjecture imposes constraints on this boundary condition.
 
Indeed, the 3d mirror of $(A_1, A_{2n-2})$ has no Coulomb branch, and we argued that it is given by a collection of free hypermultiplets. There cannot exist any 3d $\cN=4$ gauge-theoretic description of this theory, simply because the mere presence of 3d $\cN=4$ vector multiplets implies that the Coulomb branch is non-empty. If we look at the construction involving the 4d MSYM on an interval, however, we notice that the \textbf{N} boundary condition leaves 3d $\cN=4$ vector multiplet worth of unfixed field components at the boundary. If we, for example, were to impose the same boundary condition \textbf{N} at the other end of the interval, then in the 3d limit, the low-energy modes of the 4d fields would be effectively described by the true 3d $\cN=4$ vector multiplet. In other words, we would obtain a gauge theory, which as we said is undesirable. Therefore, whatever boundary condition \textbf{X} is, it should, in cooperation with \textbf{N}, completely freeze the bulk, leaving no ``trapped'' degrees of freedom in the IR. Such pairs of boundary conditions, like \textbf{N} and \textbf{X} here, were called complimentary or transversal in \cite{Dedushenko:2018aox,Dedushenko:2018tgx} (see also \cite{Dimofte:2019buf} for examples of trapped degrees of freedom).
 
The standard half-BPS Neumann and Dirichlet boundary conditions form such a transversal pair (with the proper choice of splitting for the vector multiplet scalars \cite{Gaiotto:2008sa}). Therefore, a reasonable conjecture would be to guess that \textbf{X} is simply given by the Dirichlet boundary condition, with additional $n-1$ free hypermultiplets living at the boundary. Precise description of the boundary condition following from the irregular puncture might be more complicated, but it should be equivalent to this in the 
IR. The reader might assume that \textbf{X} is precisely such a boundary condition, however what we are going to say now does not depend on its precise choice. All we need to know is that in the 3d limit, all the 4d SYM fields freeze, and there are $n-1$ free hypers somehow originating from the boundary.

Now, let us move to the $(A_1, D_{2n+1})$ case, which we will analyze using similar methods below. In the class S construction, one obtains $(A_1, D_{2n+1})$ from $(A_1, A_{2n-2})$ by adding a regular puncture at the opposite tip of the cigar, see Figure \ref{fig:Dodd5d}. The corresponding boundary condition in the 4d limit is known \cite{Chacaltana:2012zy,Gaiotto:2008ak}, -- it is simply given by the Neumann boundary condition coupled to the $T[SU(2)]$ at the boundary. This immediately tells us what happens in the 3d limit. Indeed, the $T[SU(2)]$ at the boundary interacts with the bulk through the \textbf{N} boundary conditions, and then the bulk further mediates its interaction with the other boundary. Because the (\textbf{N}, \textbf{X}) pair of boundary conditions completely freezes the bulk fields in the 3d limit, it implies that the $T[SU(2)]$ simply decouples, and the rest of the system is identical to what we studied above for $(A_1, A_{2n-2})$.  In other words, the 3d mirror of $(A_1, D_{2n+1})$ should be given by the $T[SU(2)]$ and a decoupled sector given by the 3d mirror of $(A_1, A_{2n-2})$, which, as we know now, is simply a collection of $n-1$ free hypers. This is our main proposal, which we are going to test now.
 
 \subsection{The reduction of $(A_1, D_{2n+1})$.}
As suggested above, the 3d limit of $(A_1, D_{2n+1})$ contains a decoupled sector described by $T[SU(2)]$, and the rest is identical to the 3d limit of $(A_1, A_{2n-2})$. For the $(A_1, A_{2n-2})$ part, we expect the same exact behavior: the Coulomb branch operators $u_1, \dots, u_{n-1}$ flow to composites $q_a q_a$, and the values of FI parameters are precisely the same as in the pure $(A_1, A_{2n-2})$ case. The remaining Coulomb branch operator $u_n$ has dimension $\Delta_n = 2 - \frac{2}{2n+1}$, and we expect it to flow to the dimension 2 monopole operator in the $T[SU(2)]$ (see Figure \ref{fig:Dodd} for an expected RG flow of the Coulomb branch chiral spectrum). Indeed, since the VOA reduction implies the FI parameter $\zeta= \frac{i}{\ell} \frac{1}{2n+1}$, as given in \eqref{A1DoddFI}, the mixing relation in the $T[SU(2)]$ sector becomes
 \begin{equation}
 \label{mixing_T[SU(2)]}
 2-\frac{2}{2n+1} = r = R_C + \frac{1}{2n+1} \cT,
 \end{equation}
 which suggests $R_C=2$ and $\cT=-2$ as the minimal universal way to satisfy it for all $n$. We therefore propose that $u_n$ flows to the dimension 2, charge $-2$ monopole operator in $T[SU(2)]$. Recall that the minimal operators in the chiral ring of $T[SU(2)]$ have dimension 1, therefore we again propose that they are emergent in the IR.\footnote{Also note that our analysis cannot determine the sign of $\zeta$, so we could take $\zeta\to -\zeta$, and claim that $u_n$ flows to the charge $+2$ monopole operator in 3d. Distinguishing these two cases requires a more refined analysis.}
 
To test these claims, we similarly carry out the reduction of the flavored Schur index of the $(A_1,D_{2n+1})$ theory,
 \ie
 \cI_{(A_1,D_{2n+1})}={\rm PE}\bigg[
 {(q-q^{2n+1}) \chi_{1}(z)  
 	\over 
 	(1-q)(1-q^{2n+1})	
 }
 \bigg],
 \fe
 where $\chi_1$ denotes the adjoint character for the $SU(2)$ flavor symmetry, and $z$ is a flavor fugacity related to the 3d mass $m$ via $z=q^{im}$. Taking the Cardy limit in this case, we find
 \ie
 \log \cI _{(A_1,D_{2n+1})}
 =&
 { n \over 4(2n+1)} {2\pi i\over \tau} 
 \\
 +&\sum_{j=1}^{2n} 
 \left(\log {\Gamma((j+im)/(2n +1))\over \sqrt{2\pi}}+\log {\Gamma((j-im)/(2n+1))\over \sqrt{2\pi}}
 +\log {\Gamma((j)/(2n+1))\over \sqrt{2\pi}}\right)
 +\cO(\log q).
 \fe
 Once again, the leading term is consistent with the Cardy formula \eqref{4dcardyF} using
 \ie
 c^{(A_1,D_{2n+1})}_{4d}-a^{(A_1,D_{2n+1})}_{4d}={3n\over 24(2n+1)}.
 \fe
 The finite piece determines the $S^3$ partition function of the 3d SCFT from $S^1$ reduction
 \ie
 Z_{\rm 3d}
 =\prod_{j=1}^{2n} {\Gamma\left(j\over 2n+1\right)\over \sqrt{2\pi}} {\Gamma\left(j+ im\over 2n+1\right)\Gamma\left(j- im\over 2n+1\right)\over \sqrt{2\pi}}
 %=\prod_{j=1}^{2N} {\Gamma\left(j\over 2n+1\right)\over \sqrt{2\pi}} 
 %\prod_{j=1}^{2N} {1\over 2 \sinh  {\pi(j+i \sigma)\over 2n+1} }
 = \frac1{2\sin\frac{\pi}{2n+1}} { \sinh  {\pi  m \over (2n+1)}\over \sinh \pi m } \prod_{j=2}^{2n-1} {\Gamma\left(j\over 2n+1\right)\over \sqrt{2\pi}} .
 \fe
 This agrees with the factorization
 \ie
 Z_{\rm 3d}= Z_{\rm SQED_2}(\zeta,m)\prod_{a=1}^{n-1} Z_{\eta}(\zeta_a),\quad \text{where } Z_{\rm SQED_2}(\zeta,m)=\frac1{2\sinh\pi\zeta} { \sin  {\pi  m  \zeta}\over \sinh \pi m },
 \fe
 where the FI parameters $\zeta_a$ are given in \eqref{A1AevenFI}, and $Z_{\rm SQED_2}(\zeta,m)$ denotes the $S^3$ partition function for SQED$_2$ \cite{Benvenuti:2011ga} with an arbitrary  mass parameter $m$ and the FI parameter $\zeta$ from  \eqref{A1DoddFI}. To avoid clutter in the final expression, we put $\ell=1$, which can be easily restored by dimensional analysis.
 
 We see a complete agreement with the proposal that the 3d limit is given by the $T[SU(2)]$ and $n-1$ decoupled twisted hypers. To complete the story, we also test the 3d limit in the Coulomb branch lens space index, just like we did in the $(A_1, A_{2n-2})$ case. The lens space Coulomb index for the $(A_1, D_{2n+1})$ theories was computed in \cite{Fredrickson:2017yka} as well,
 \begin{align}
 \cI_{n,p}(t,\lambda) &= \frac{1}{\prod_{k=1}^n \left( 1-t^{\frac{2k-1}{2n+1}} \right)\left( 1-t^{\frac{2n+2-2k}{2n+1}} \right)}\cr 
 &+ \sum_{i=1}^n \frac{t^{p\mu^{(1)}_i} + t^{p\mu^{(2)}_{i}}}{\prod_{k=1}^i\left(1-t^{\frac{2n+2k}{2n+1}}\right)\left(1-t^{-\frac{2k-1}{2n+1}}\right)\prod_{k=i+1}^n \left( 1-t^{\frac{2k-1}{2n+1}}\right)\left(1-t^{\frac{2n+2-2k}{2n+1}} \right)},
 \end{align}
 where
 \begin{equation}
 \mu^{(1)}_i=\frac{i(i+1)}{2(2n+1)}-\frac{i}{2n+1}\cdot\frac{\lambda}{p},\quad \mu^{(2)}_i=\frac{(i-1)i}{2(2n+1)} + \frac{i}{2n+1}\cdot \frac{\lambda}{p},
 \end{equation}
 and $\lambda\in \{0,1,\dots ,p\}$ denotes the holonomy around the 1-cycle of the lens space for the flavor $SU(2)$ symmetry of $(A_1, D_{2n+1})$. In order to take the 3d limit, we ought to turn off this holonomy, $\lambda=0$, and send $p\to\infty$, again assuming that $|t|<1$. Only two terms survive this limit,
 \begin{align}
 \lim_{p\to\infty} \cI_{n,p}(t,0)&=\frac{1}{\prod_{k=1}^n \left( 1-t^{\frac{2k-1}{2n+1}} \right)\left( 1-t^{\frac{2n+2-2k}{2n+1}} \right)}\cr 
 &+ \frac{1}{\left(1-t^{\frac{2n+2}{2n+1}}\right)\left(1-t^{-\frac{1}{2n+1}}\right)\prod_{k=2}^n \left( 1-t^{\frac{2k-1}{2n+1}}\right)\left(1-t^{\frac{2n+2-2k}{2n+1}} \right)}\cr
 &= \frac{1+t}{\left(1-t^{\frac{2n}{2n+1}}\right)\left(1-t^{\frac{2n+2}{2n+1}} \right)} \prod_{j=2}^{2n-1}\frac1{1-t^{\frac{j}{2n+1}}},
 \end{align}
 where we have massaged the last expression into a very suggestive form. The product part coincides with \eqref{CoulIndLimit}, and thus counts the chiral spectrum of free twisted hypermultiplets. The factor
 \begin{equation}
 \label{index_of_TSU2}
 \frac{1+t}{\left(1-t^{\frac{2n}{2n+1}}\right)\left(1-t^{\frac{2n+2}{2n+1}} \right)},
 \end{equation}
 therefore, must be counting the Coulomb branch spectrum of $T[SU(2)]$, if our proposal is correct. The Coulomb branch chiral ring of $T[SU(2)]$  is generated by three operators $X, Y, Z$, whose dimensions and $R_C$-charges are all $1$, and the topological charges are $+1$, $-1$, and $0$, respectively. To account for the mixing \eqref{mixing_T[SU(2)]}, we compute their r-charges,
 \begin{align}
 \label{mixed_charge_TSU2}
 r(X) = 1 + \frac1{2n+1}=\frac{2n+2}{2n+1},\quad r(Y) = 1 - \frac1{2n+1}=\frac{2n}{2n+1},\quad r(Z)=1.
 \end{align}
Observe that these are precisely the powers that appear in \eqref{index_of_TSU2}. Also, the generators satisfy $XY=Z^2$, so general chiral operators can be written as $P_0(X,Y) + Z P_1(X,Y)$, where $P_{0,1}$ are arbitrary polynomials. Counting such operators weighted by $t^r$ indeed gives \eqref{index_of_TSU2}, which is thus the Coulomb branch index\footnote{Of course it is also the Higgs branch index, since $T[SU(2)]$ is self-mirror.} of $T[SU(2)]$ with the unusual weight factor $t^r$, where the $r$ charges are given by \eqref{mixed_charge_TSU2}. This confirms our proposal again. Notice that the operators $X, Y, Z$ do not come directly from the 4d Coulomb branch operators, so they must emerge in the 3d limit from line operators wrapping the shrinking 1-cycle.

\section{Lagrangian theories}\label{sec:LagrangianEx}
Let us now focus on the class of examples coming from Lagrangian 4d $\cN=2$ theories. The sort of mixing discussed before does not occur in such cases, which slightly simplifies the story, as we do not have to worry about the FI terms. More importantly, a large set of Lagrangian tools makes them more amenable to explicit analysis.
\subsection{Matrix model reduction}
The flavored $S^3\times S^1$ partition function has a matrix model description that follows from the localization (see, \emph{e.g.}, \cite{Romelsberger:2007ec,Gadde:2011ik,Aharony:2013dha,Assel:2014paa}), -- we will use the form of the answer employed in \cite{Dedushenko:2019yiw,Pan:2019bor}. For concreteness, let us focus on theories built from vectormultiplets and full hypermultiplets, and write the answer in terms of Jacobi theta functions,
\begin{equation}
\label{matrmodZ}
Z = \frac{q^{-\frac14 \sum \langle w_f, a_f\rangle^2}}{|\cW|}\int_0^{2\pi}\left[\frac{\dd a}{2\pi}\right]^r \eta(\tau)^{3r-|G| + |\cR|} \frac{\prod_{\alpha\in\Delta\setminus 0} \theta_1(\langle\alpha,a\rangle/2\pi; \tau)}{\prod_{(w,w_f)\in\cR} \theta_4((\langle w,a\rangle+ \langle w_f, a_f\rangle)/2\pi; \tau)},
\end{equation}
where $w$ and $w_f$ denote gauge and flavor weights of the matter multiplets respectively. The 3d, or high-temperature, limit of this partition function has been analyzed in great details and in full generality in \cite{Ardehali:2015bla}, and the readers should consult that reference for more details. In the simplest case of theories obeying $c_{\rm 4d}> a_{\rm 4d}$, it is enough to perform a somewhat more simplistic analysis to get the correct result. First, it is necessary to write the integration region as $a\in (-\pi, \pi)^r$, rather than $(0,2\pi)^r$, -- this change does not affect the exact answer due to periodicity of the integrand, yet is important for proper asymptotic analysis, as the leading contribution comes from the region close to $a=0$. Next, we rescale the integration variable $a$ and the flavor fugacity $a_f$ according to
\begin{equation}
a = \beta\sigma,\quad a_f = \beta m_f.
\end{equation}
The integration now ranges over $\sigma\in \left(-\frac{\pi}{\beta}, \frac{\pi}{\beta} \right)$. Now we take the $\beta\to0$ limit and approximate the integrand by its $\beta\to0$ asymptotics. While for the eta-function it is enough to write $\eta(\tau)\approx \frac1{\sqrt{-i\tau}}\widetilde{q}^{\frac1{24}}$, with $\widetilde{q}=e^{-2\pi i/\tau}=e^{-4\pi^2/\beta}$, for theta function it is useful to perform the modular S transform and use the product formula, which gives:
\begin{align}
\theta_1(z;\tau) &= \frac{i}{\sqrt{-i\tau}} e^{-\frac{\pi}{\tau}iz^2} 2\sin\left(\pi\frac{z}{\tau} \right)\widetilde{q}^{\frac18} \left[\left(1-e^{2\pi i\frac{z}{\tau}}\widetilde{q} \right)\left(1-e^{-2\pi i\frac{z}{\tau}} \widetilde{q} \right) \right] \left[\left(1-e^{2\pi i\frac{z}{\tau}}\widetilde{q}^2 \right)\left(1-e^{-2\pi i\frac{z}{\tau}} \widetilde{q}^2 \right) \right]\dots \cr
\theta_4(z;\tau) &= \frac1{\sqrt{-i\tau}} e^{-\frac{\pi}{\tau}iz^2} 2\cos\left(\pi\frac{z}{\tau} \right)\widetilde{q}^{\frac18} \left[\left(1+e^{2\pi i\frac{z}{\tau}}\widetilde{q} \right)\left(1+e^{-2\pi i\frac{z}{\tau}} \widetilde{q} \right) \right] \left[\left(1+e^{2\pi i\frac{z}{\tau}}\widetilde{q}^2 \right)\left(1+e^{-2\pi i\frac{z}{\tau}} \widetilde{q}^2 \right) \right]\dots
\end{align}
When $z$ is small, the factors in square brackets can be neglected in the $\tau\to +0i$ limit, but they start contribute one by one as we increase $z$ up to $1$, $2$ etc. In general, as we take the $\tau\to+0i$ limit, one might encounter several saddle points in the integral \eqref{matrmodZ} as was found in \cite{Ardehali:2015bla}. Again, the simplest situation is when only the saddle near $a=0$ contributes, in which case we approximate theta functions near $z=0$, and all factors in the square brackets can be dropped:
\begin{align}
\theta_1(z;\tau) \approx \frac{i}{\sqrt{-i\tau}} e^{-\frac{\pi}{\tau}iz^2} 2\sin\left(\pi\frac{z}{\tau} \right)\widetilde{q}^{\frac18},\quad 
\theta_4(z;\tau) \approx \frac1{\sqrt{-i\tau}} e^{-\frac{\pi}{\tau}iz^2} 2\cos\left(\pi\frac{z}{\tau} \right)\widetilde{q}^{\frac18}.
\end{align}
With these approximations, one finds the $\tau\to+0i$ asymptotics:
\begin{equation}
\label{matri_mod_red}
Z \approx \widetilde{q}^{\frac1{12}(\dim(G) - \dim_\C(\cR))} \frac1{|\cW|} \int_{\mathfrak t} \dd^r\sigma \frac{\prod_{\alpha\in\Delta\setminus 0}2\sinh\left(\langle\alpha,\sigma\rangle \right)}{\prod_{(w,w_f)\in\cR} 2\cosh\left( \pi\left(\langle w,\sigma\rangle + \langle w_f,m_f\rangle \right) \right)},
\end{equation}
which is the well-known Kapustin-Willett-Yaakov matrix model in 3d \cite{Kapustin:2009kz}, with the divergent factor implying $c_{\rm 4d}-a_{\rm 4d}=\frac1{24}(\dim_\C(\cR) - \dim(G))$. In particular, theories with enough matter satisfy $c_{\rm 4d}>a_{\rm 4d}$.

One can also repeat this exercise and reduce the matrix model coupled to symplectic bosons on the two-torus, as derived in \cite{Dedushenko:2019yiw,Pan:2019bor}, to a matrix model coupled to quantum mechanics on the circle found in \cite{Dedushenko:2016jxl}. In doing this, we integrate out all non-zero Kaluza-Klein modes on the torus, while only keeping the zero modes of symplectic bosons unintegrated. We then take the $\tau\to +0i$ limit, similar to the above discussion. For theories admitting the standard high-temperature behavior governed by the Di Pietro-Komargodski formula, this works straightforwardly: the Cardy-like term is precisely as in \eqref{matri_mod_red}, and the correlators behave well in the $\tau\to+0i$ limit.
\subsection{Testing the W-algebra reduction: $SU(3)$ SQCD}
Since the $SU(2)$ SQCD has already been analyzed, here we turn to the next simplest example -- the $SU(3)$ SQCD, which is much more involved because the corresponding VOA is a W-algebra \cite{Beem:2013sza}. In fact, for  $SU(N)$, $N_f=2N$ gauge theory with arbitrary $N\geq 3$, the W-algebra has been conjectured in \cite{Beem:2013sza}  to be generated by the $\mathfrak{su}(N_f)_{-N}\oplus \mathfrak{u}(1)_{-N}$ affine currents,
\begin{align}
J_i{}^j(z)J_k{}^l(0) &\sim -\frac{N\left(\delta_i^l\delta_k^j - \frac{1}{N_f}\delta_i^j \delta_k^l\right)}{z^2} + \frac{\delta_i^lJ_k{}^j (0) - \delta_k^j J_i{}^l(0)}{z},\cr
J(z)J(0) &\sim -\frac{2N^2}{z^2},
\end{align}
together with an additional pair of strong generators at level $L_0=\frac{N}{2}$.  These are Virasoro primaries corresponding to the baryonic chiral ring generators given by the antisymmetric $SU(N_f)$-tensors with $\frac{N_f}{2}=N$ flavor indices each,
\begin{align}
\label{W_b}
b_{i_1 i_2\dots i_N} = \varepsilon^{\alpha_1\alpha_2\dots\alpha_N}q_{\alpha_1 i_1}q_{\alpha_2 i_2}\dots q_{\alpha_{N}i_N},\cr
\widetilde{b}^{i_1 i_2\dots i_N} = \varepsilon_{\alpha_1\alpha_2\dots\alpha_N}\widetilde{q}^{\alpha_1 i_1}\widetilde{q}^{\alpha_2 i_2}\dots \widetilde{q}^{\alpha_{N}i_N},\cr
\end{align}
where we adopt the notations of \cite{Beem:2013sza}. In particular, $q$, $\widetilde{q}$ are symplectic bosons, $\varepsilon$ is the $SU(N)$-invariant tensor, and $\alpha_i$ are gauge indices. In these notations, the affine currents are
\begin{align}
\label{affine_J}
J_i{}^j &= q_{\alpha i}\widetilde{q}^{\alpha j} -\frac{1}{N_f}\delta_i^j q_{\alpha k}\widetilde{q}^{\alpha k},\cr
J&=q_{\alpha k}\widetilde{q}^{\alpha k},
\end{align}
and the only non-obvious OPE one should specify is between $b$ and $\widetilde{b}$. For $N=3$, it is given in \cite{Beem:2013sza}:
\begin{equation}
b_{i_1i_2i_3}(z)\widetilde{b}^{j_1j_2j_3}(0) \sim \frac{36\delta_{[i_1}^{[j_1}\delta_{i_2}^{j_2}\delta_{i_3]}^{j_3]}}{z^3} - \frac{36\delta_{[i_1}^{[j_1}\delta_{i_2}^{j_2}J_{i_3]}{}^{j_3]}(0)}{z^2} + \frac{18\delta_{[i_1}^{[j_1}\left(J_{i_2}{}^{j_2}J_{i_3]}{}^{j_3]}\right)(0)-18\delta_{[i_1}^{[j_1}\delta_{i_2}^{j_2}\partial J_{i_3]}{}^{j_3]}(0)}{z}.
\end{equation}
The stress-energy tensor coincides with the one provided by the Sugawara construction,
\begin{equation}
\label{SU3_T}
T=\frac1{N_f}\left( (J_i{}^j J_j{}^i) - \frac1{N_f}(JJ) \right),
\end{equation}
with the two-dimensional central charge
\begin{equation}
c_{\rm 2d}=2-4N^2.
\end{equation}
We now want to study the dimensional reduction for this model from various points of view, verifying our general results and exemplifying various further subtleties that appear in this problem.

\subsubsection{Analysis in 3d}
The 3d reduction is the $SU(N)$ gauge theory with $N_f=2N$ fundamental hypers, and as usual, we apply the techniques of \cite{Dedushenko:2016jxl}. Now it is convenient to work with full hypers, and their 1d relatives $Q_{\alpha i}$ and $\widetilde{Q}^{\alpha i}$ have the following propagator at the fixed Coulomb parameter $\sigma$,
\begin{align}
\label{SQCD_prop}
\langle Q_{\alpha i}(\varphi_1) \tilde{Q}^{\beta j}(\varphi_2)\rangle_\sigma &= \delta_i^j (G_\sigma)_\alpha{}^\beta(\varphi_1-\varphi_2),\cr
G_\sigma(\varphi) &= -\frac{{\rm sgn}(\varphi)+\tanh(\pi\sigma)}{8\pi \ell}e^{-\sigma\varphi},\quad G_\sigma(0)=-\frac{\tanh(\pi\sigma)}{8\pi \ell}.\cr
\end{align}
The generators $J_i{}^j$, $J$, $b$, and $\widetilde{b}$ have the same expressions in 1d as their 2d counterparts \eqref{affine_J} and \eqref{W_b}, with $q$, $\widetilde{q}$ replaced by $Q$, $\widetilde{Q}$. Using the propagator \eqref{SQCD_prop}, we find for general $N$:
\begin{align}
\label{JJ_from_3d}
\langle J_i^j(\varphi) J_k^l(0)\rangle_\sigma&=-\frac1{(8\pi\ell)^2}\left(\delta_i^l\delta_k^j - \frac1{N_f}\delta_i^j \delta_k^l\right)\trace_\cR \left[\frac1{\cosh^2(\pi\sigma)} \right]\,,\cr
\langle J(\varphi) J(0)\rangle_\sigma &= -\frac{N_f}{(8\pi\ell)^2}\trace_\cR \left[\frac1{\cosh^2(\pi\sigma)} \right] + \frac{N_f^2}{(8\pi\ell)^2} \left[\trace_\cR \tanh(\pi\sigma) \right]^2\,.\cr
\end{align}
We next have to compute integrals over sigma, and in particular the partition function is
\begin{align}
Z=\int_{\R^2} \dd^r\vec\sigma \frac{\prod_{\alpha\in\Delta\setminus 0}2\sinh(\pi \langle\alpha, \sigma^\vee\rangle)}{\left(\prod_{w\in\cR}2\cosh(\pi\langle w,\sigma^\vee\rangle)\right)^{N_f}},
\end{align}
where $\vec\sigma=(\sigma_1, \sigma_2, \dots, \sigma_r)$ and $\sigma^\vee = \sum_{i}\sigma_i \alpha_i^\vee$, with $\alpha_i^\vee$ the coroots (this is simply a convenient choice of normalization of $\sigma$). For general $N$, such integrals are complicated and can be studied using the techniques developed in \cite{Russo:2014bda,Tierz:2016zcn,Tierz:2016fom,Tierz:2018fsn,Garcia-Garcia:2019uve}. Focusing on the case $N=3$, we obtain
\begin{align}
\left\langle \trace_\cR \left[\frac1{\cosh^2(\pi\sigma)} \right] \right\rangle = \frac{46848-4725 \pi^2}{35 \left(448-45 \pi^2\right)}, \quad \quad
\left\langle \left[\trace_\cR \tanh(\pi\sigma) \right]^2 \right\rangle =\frac{8 \left(1575 \pi ^2-15544\right)}{35 \left(448-45 \pi ^2\right)}\,,
\end{align}
thus giving
\begin{align}
\langle J_i^j(\varphi) J_k^l(0)\rangle&=-\frac1{(8\pi\ell)^2}\left(\delta_i^l\delta_k^j - \frac1{6}\delta_i^j \delta_k^l\right)\frac{46848-4725 \pi^2}{35 \left(448-45 \pi^2\right)}\,,\cr
\langle J(\varphi) J(0)\rangle &= -\frac{1}{(8\pi\ell)^2}\frac{18 \left(7552-765 \pi^2\right)}{448-45 \pi^2}\,.\cr
\end{align}
It is also straightforward to compute for $\varphi_1>\varphi_2>\varphi_3$,
\begin{align}
\langle J_i^j(\varphi_1) J_k^l(\varphi_2) J_m^n(\varphi_3)\rangle &=\frac1{(8\pi\ell)^3}(\delta_i^l\delta_m^j\delta_k^n-\delta_i^n\delta_k^j\delta_m^l - \text{traces})\left\langle \trace_\cR \left[\frac1{\cosh^2(\pi\sigma)} \right] \right\rangle\,,\cr
\langle J(\varphi_1) J(\varphi_2) J(\varphi_3)\rangle&=0\,,
\end{align}
and thus identify star products in the subalgebra of currents as
\begin{align}
\label{SU3_3dAlg}
J_i^j \star J_k^l &= :J_i^j J_k^l: - \frac{1}{8\pi\ell}(\delta_i^l J_k^j - \delta_k^j J_i^l) +\frac1{(8\pi\ell)^2}\left(\delta_i^l\delta_k^j - \frac1{N_f}\delta_i^j \delta_k^l\right) \mu_1\,,\cr
J \star J &= :JJ: +\frac1{(8\pi\ell)^2}\mu_2\,,
\end{align}
where for $N=3$,
\begin{equation}
\label{SU3_mu12_values}
\mu_1=-\frac{46848-4725 \pi^2}{35 \left(448-45 \pi^2\right)}\,,\quad \mu_2=-\frac{18 \left(7552-765 \pi^2\right)}{448-45 \pi^2}\,.
\end{equation}
Here, as usual, $:AB:$ is defined as the highest-dimension operator appearing in $A\star B$, made orthogonal to all the lower-dimension operators via the Gram-Schmidt procedure. In the above expression, this means that $:J_i{}^j J_k{}^l:$ and $:JJ:$ are necessarily orthogonal to the identity and the dimension-1 operators (currents), but they might not be orthogonal to each other, \emph{i.e.} there is still some residual mixing between $:J_i{}^j J_k{}^l:$ and $:JJ:$. Also notice that in our case there exist dimension-$\frac32$ operators $b$ and $\widetilde{b}$, but $:J_i{}^j J_k{}^l:$ and $:JJ:$ obviously cannot mix with them.

\subsubsection{Analysis in 4d}
Let us now verify that the same answers follow from the high-temperature limit of torus correlators. Like in section \ref{sec:affineVOA}, it is straightforward to write the currents two-point functions by matching the pole structure in the OPE,
\begin{align}
\langle J_i^j(z_1) J_k^l(z_2)\rangle &= -\frac{N}{\ell^2}\left(\delta_i^l\delta_k^j - \frac1{N_f}\delta_i^j \delta_k^l\right)\left(\frac1{(2\pi)^2}\wp\left(\frac{z_1-z_2}{2\pi};\tau \right) + e_1(\tau) \right),\cr
\langle J(z_1) J(z_2)\rangle &= -\frac{2N^2}{\ell^2}\left(\frac1{(2\pi)^2}\wp\left(\frac{z_1-z_2}{2\pi};\tau \right) + e_2(\tau) \right),
\end{align}
with the only difference that now we have two $z$-independent functions $e_{1,2}(\tau)$, since the underlying Lie algebra is not simple and has precisely two direct summands. They are of course given by the one-point functions of the normal products of currents,
\begin{align}
\label{JJ_1ptf}
\langle (J_i^j J_k^l)\rangle &= -\frac{N}{\ell^2}\left(\delta_i^l\delta_k^j - \frac1{N_f}\delta_i^j \delta_k^l\right)e_1(\tau),\cr
\langle (JJ)\rangle &= -\frac{2N^2}{\ell^2}e_2(\tau).
\end{align}
We know that the Weierstrass function behaves as $\wp\sim -\frac{pi^2}{3\tau^2}$ in the $\tau\to0$ limit, so it only remains to find the high-temperature asymptotics of $e_{1,2}(\tau)$. The natural expectation, which we will confirm below, is that the behave as
\begin{equation}
\label{expect_e12}
e_1(\tau) \sim \frac{A}{\tau^2},\quad e_2(\tau)\sim\frac{B}{\tau^2},
\end{equation}
Implying the following limits
\begin{align}
\lim_{\tau\to 0} \tau^2 \langle J_i^j(z_1) J_k^l(z_2)\rangle &= \hbar^2 4N\left(\delta_i^l\delta_k^j - \frac1{N_f}\delta_i^j \delta_k^l\right)\left( -\frac1{12} + A \right),\cr
\lim_{\tau\to 0} \tau^2 \langle J(z_1) J(z_2)\rangle &= \hbar^2 8N^2\left( -\frac1{12} + B \right),
\end{align}
which result in the same 1d algebra as given in \eqref{SU3_3dAlg}, with the relation between $A, B$ and $\mu_1, \mu_2$ as follows,
\begin{align}
\label{AB_from_mu}
A=\frac1{12} + \frac{\mu_1}{4N},\quad B=\frac1{12}+\frac{\mu_2}{8N^2}.
\end{align}
These values of $A,B$ also agree with the Cardy behavior, as one can easily check. Indeed, the one-point function of the stress tensor \eqref{SU3_T} follows from the one point functions \eqref{JJ_1ptf}, and is given by
\begin{equation}
\langle T\rangle = -\frac{(N_f^2-1)e_1(\tau)}{2\ell^2} + \frac{e_2(\tau)}{2\ell^2}.
\end{equation}
Using the asymptotics \eqref{expect_e12}, with $A, B$ given in \eqref{AB_from_mu}, and $\mu_1, \mu_2$ taking values \eqref{SU3_mu12_values}, we find
\begin{equation}
\langle T\rangle \sim \frac{5/6}{\ell^2\tau^2}.
\end{equation}
On the other hand,  the torus partition function $Z$ behaves in accordance with \eqref{Cardy_Z},
\begin{equation}
Z\sim \widetilde{q}^{\frac{\dim(G) -N_f\dim_\C(\cR) }{12}}= e^{\frac{i\pi}{6\tau}(N^2+1)},
\end{equation}
indeed implying the same behavior of $\langle T\rangle$,
\begin{equation}
\langle T\rangle = -\frac{1}{2\pi i \ell^2} \frac{\dd}{\dd\tau} \log(Z) \sim \frac{N^2+1}{12\ell^2\tau^2} = \frac{5/6}{\ell^2\tau^2}.
\end{equation}
We also confirm the behavior \eqref{expect_e12} of $e_1(\tau)$ and $e_2(\tau)$, with $A,B$ given in \eqref{AB_from_mu}, from the chiral algebra computation in Appendix \ref{app:JJ_1pf}. 

\subsubsection{Star products from the flavored Schur index}
Another method to extract the $\tau\to0$ limits of correlators of currents is by looking at the $\tau\to0$ limit of the flavored Schur index. It is somewhat redundant in the present case since it obviously gives the same results, but will be the only technique available to us in later sections. The flavored Schur index gives the flavored vacuum character of the VOA,
\begin{equation}
I(\tau, a_f) = \trace_V q^{L_0-\frac{c_{\rm 2d}}{24}} u_f^{J^f_0},
\end{equation}
where $u_f=e^{i a_f}$ is the flavor fugacity, and $J^f_0 = \oint \frac{\dd z}{2\pi} J^f(z)$ is the charge of the affine current $J^f(z)$, with the integral going over the spacial $S^1\subset S^3$. In the high-temperature limit, as was previously mentioned, we scale $a_f= \beta m_f = -2\pi i \tau m_f$, and $m_f$ becomes the 3d mass. By taking derivatives with respect to $m_f$, we can compute integrated correlators of currents in 1d. Because the 1d limit of the two-point function of currents is a constant (which is simply the constant term of $J^f \star J^f$), this allows to compute the 1d two-point function of two arbitrary currents. Namely,
\begin{equation}
\frac{1}{I} \frac{\partial^2 I}{\partial m_f^2}\Big|_{m_f=0} = -(2\pi)^2 \tau^2 \oint \frac{\dd z_1}{2\pi} \oint \frac{\dd z_2}{2\pi} \langle J^f(z_1) J^f(z_2)\rangle.
\end{equation}
Taking the $\tau\to0$ limit, the $z$-dependence of the two-point function drops out, the integrals become trivial, and we can simply write,
\begin{equation}
\langle J_1 \star J_2\rangle = -\lim_{\tau\to +0i} \frac{1}{(2\pi)^2 I(\tau, m)} \frac{\partial^2 I(\tau, m)}{\partial m_1 \partial m_2} \Big|_{m=0},
\end{equation} 
where we generalized to the case of several currents $J_i$ with the associated masses denoted $m_i$. In the next Section, this will help to extract star products directly from the flavored Schur index.

\section{More on Argyres-Douglas theories}\label{sec:ADMore}

 The general construction of Argyres-Douglas theories from twisted compactification of $(2,0)$ theory of ADE type $\mf j$ on a Riemann surface $\cC$ with irregular and regular punctures were carried out in \cite{Wang:2015mra,Wang:2018gvb} extending previous results for the $A$-type case \cite{Gaiotto:2009hg,Nanopoulos:2009uw,Bonelli:2011aa,Xie:2012hs,Xie:2013jc}. The requirement of superconformal symmetry fixes $\cC$ to be a sphere (with holomorphic coordinate $z$) and the irregular puncture at $z=0$ to be described by a singularity (Hitchin pole) of the Higgs field
 \ie
 \Phi={T\over z^{2+{\kappa\over b}}}+\dots 
 \fe
 where $T$ is an element in the Cartan subalgebra of $\mf j$, $b$ is selected from a set of positive integers fixed by $\mf j$, and $\kappa \in \bZ$ satisfies $\kappa >-b$. The construction allows another regular puncture at $z=\infty$ labeled by $Y$, which is a Young-tableau that captures the Hitchin partition for the classical Lie algebras $\mf j$, and more generally given by Bala-Carter labels for the exceptional cases. The SCFT constructed by such a pair of irregular and regular punctures in the 6d $(2,0)$ theory is denoted $(J^b[\kappa],Y)$, and the one that is produced by the irregular puncture alone is named $J^b[\kappa]$ \cite{Wang:2015mra}.\footnote{Note that it is common for a given 4d $\cN=2$ SCFT to have multiple class S constructions and this gives rise to identifications among the labels $(J^b[\kappa],Y)$.}

 The 2d chiral algebra for the $(J^b[\kappa],Y)$ theory is a W-algebra \cite{Song:2017oew} of the type\footnote{In general there are certain constraints on $(\mf j, b,\kappa)$ for this statement to be true  \cite{Song:2017oew}. Such $(\mf{j}, b, \kappa)$ exist for all the cases we consider in this paper.
 }
 \ie
 W^k(\mf{j},Y),\quad k=-h^\vee+{b\over \kappa +b}
 \label{ADW}
 \fe
 where $h^\vee$ denotes the dual Coxeter number of $\mf j$ (see also Table 5 in \cite{Xie:2019zlb}). Such W-algebra arises from the quantum-Drinfeld-Sokolov (qDS) reduction of the affine Kac-Moody (AKM) algebra $V_k(\mf{j})$, which corresponds to the special case of \eqref{ADW} when $Y=F$ labels a full (principal) regular puncture.
 
 The Higgs branch of the SCFT $(J^b[\kappa],Y)$ is identified with the associated variety of the W-algebra, and given by the intersection of the Slodowy slice $S_Y$ transverse to the coadjoint nilpotent orbit $\cO_Y$ with another nilpotent orbit,
 \ie
 \cM_{\rm HB}=\begin{cases}
 	\cN_{\mf j}\cap S_Y & ~{\rm if~} \kappa>0,
 	\\
 	X_M\cap S_Y & ~{\rm if~}  \kappa<0,
 \end{cases}
 \fe
 where $\cN_{\mf j}$ denotes the nilpotent cone and $X_M$ is the closure of certain nilpotent orbits of $\mf j$ that depends on $k$ \cite{Beem:2017ooy,Song:2017oew}. 
 
 We  focus on the case $\mf j= {\mf{sl}}_n$ in this paper. In previous sections (section~\ref{sec:A1Dodd} and~\ref{sec:checkFTH}), we have studied the special cases where the 2d chiral algebra is either a Virasoro algebra (for $(A_1,A_{2n-2})$ theories which correspond to $A_{1}^2[2n-1]$ above) or an AKM algebra (for $(A_1,D_{2n+1})$ theories correspond to $(A_{1}^2[2n-1],[1,1])$). Below we extend the analysis to more general Argyres-Douglas SCFTs that realize nontrivial W-algebras.

 \subsection{$(A_1,A_{2n-1})$ AD theories}
 The $(A_1,A_{2n-1})$ SCFT  has an $n-1$ complex dimensional Coulomb branch and a two complex dimensional Higgs branch, with $U(1)$ global symmetry (enhanced to $SU(2)$ for the $n=2$ case). The
 conformal central charges are
 \ie
 a_{4d}={12n^2-5n-5\over 24(n+1)},\quad c_{4d}={3n^2-n-1\over 6(n+1)}.
 \fe
 This is a special case of the $(J^b[\kappa],Y)$ theories in \cite{Wang:2015mra} as $(A_{n-1}^n[1], [n-1,1])$. Their Higgs branch 
 is described by the intersection of the Slodowy slice of $\mf{sl}_{n}$, transverse to the subregular nilpotent orbit $[1,n-1]$, with the nilpotent cone $\cN_{\mf{sl}_n} $,
 \ie
 \cM_{\rm HB}=\overline\cN_{\mf{sl}_n} \cap \cS_{[n-1,1]}
 =
 {\bC^2/\bZ_n},
 \fe
 which is equivalent to the $A_{n-1}$ singularity. The chiral algebra here is \cite{Song:2017oew}
 \ie
 \cW^{-n^2\over n+1}(A_{n-1},[n-1,1]),
 \fe
  which contains a $U(1)$ current subalgebra that descends from the flavor symmetry multiplet in the 4d SCFT.

 The 3d SCFT from $S^1$ reduction has the following quiver description,
 \ie
 &\xymatrix{
 	\boxed{U(1)} \ar@{-}[r]_-{~q_1,\tilde q_1} & U(1) \ar@{-}[r]_-{~q_2,\tilde q_2} & \cdots \ar@{-}[r]_-{~q_{n-1},\tilde q_{n-1}} & U(1)\ar@{-}[r]_-{~q_n,\tilde q_n} & \boxed{U(1)}\, ,
 }
\label{A1Aoddquiver}
 \fe
 with $n-1$ $U(1)$ gauge nodes. The mirror quiver is simply that of SQED$_n$,
 \ie
 &\xymatrix{
 	\boxed{SU(n)} \ar@{-}[r]_-{~p_a,\tilde p_a} & U(1)  .
 }
 \fe
 In the first description, we denote the $n$ hypermultiplets by $(q_a,\tilde q_a)$  with  $a=1,2,\dots,n$, which are subjected to the D-term relations
 \ie
 q_a \tilde q_a =q_b \tilde q_b
 \fe
 for all  $a$ and $b$. The Higgs branch chiral ring is generated by
 \ie
 J=q_1 \tilde q_1,~X=q_1 q_2 \dots q_n,~Y=\tilde q_1 \tilde q_2 \dots  \tilde q_n,
 \fe
 subject to the ring relation
 \ie
 XY=J^n.
 \fe
 In particular $J$ is the moment map operator associated with the $U(1)$ flavor symmetry.
 
 The spectrum of Coulomb branch chiral primaries in the $(A_1,A_{2n-1})$ SCFT is given by 
 \ie
\Delta =r= 1+{a\over n+1}~{\rm for~} a=1,2,\dots, n-1. 
\label{A1AoddCS}
 \fe
 The fractional $U(1)_r$ charge indicates that upon $S^1$ reduction, the 3d $SU(2)_C$ R-symmetry Cartan $R_C$ must be given by a combination of the 4d $U(1)_r$ generator and emergent Coulomb branch topological $U(1)$ symmetries. The 4d Coulomb branch chiral primaries map to monopole operators in the quiver gauge theory \eqref{A1Aoddquiver}, while in the mirror SQED$_n$ description with the hypermuliplets denoted by $p_a,\tilde p^a$ for $a=1,2,\dots,n$, the relevant 3d operators are simply built from gauge invariant combinations of $p_a,\tilde p^a$.
 
  The precise R-symmetry mixing was worked out in \cite{Buican:2015hsa}, which we rewrite in a different way as
   \ie
  r=R_c-\sum_{a=1}^n {n+1-2a\over 2(n+1)}h_a,
  \label{A1AoddRmix}
  \fe
  where $h_a$ are Cartan generators of the the enhanced $SU(n)$ symmetry (manifest in the SQED$_n$ description) such that the charges of the hypermultiplets  in SQED$_n$  are normalized as 
  \ie
  h_a( p_b)=\delta_{ab} p_b,\quad h_a( \tilde p_b)=-\delta_{ab} \tilde p_b.
  \fe
Consequently, the 3d chiral primaries corresponding to the 4d Coulomb branch operators \eqref{A1AoddCS} are $p_1 \tilde p_{a+1}$ (up to mixing with operators with the same $U(1)_r$ quantum number).
    
 The FI parameters in the quiver \eqref{A1Aoddquiver} that capture the R-symmetry mixing  are 
  \ie
  \zeta_a =-{i\over \ell} {1\over n+1} ,\quad a=1,2,\dots,n-1,
  \label{A1AoddFI}
  \fe
  or equivalently in the mirror description (see Appendix~\ref{app:3dPFid} for the mirror-map) the masses are
    \ie
  m_a ={i\over \ell}{n+1-2a\over 2(n+1)}. 
  \label{A1AoddFIm}
  \fe
In the following sections, we verify that with these FI (mass) parameters, the reduction of the 4d flavored Schur index gives directly the 3d sphere partition function. Consequently, the subsector of the full TQM involving the $U(1)$ moment map operator $J$ agrees with the Cardy limit of the corresponding sector of the W-algebra.

 \subsubsection{Analysis in 4d}
 The flavored Schur index for the $(A_1, A_{2n-1})$ theory is given by
 \ie
 \cI_{(A_1,A_{2n-1})}={\rm PE}\bigg[
 {q(1+q)(1-q^{n-1})+(a+a^{-1})   q^{n \over 2}(1-q^2)
 	\over 
 	(1-q)(1-q^{n+1})	
 }
 \bigg].
 \fe
 For $n=2$, we have further simplification,
 \ie
 \cI_{(A_1,A_{3})}={\rm PE}\bigg[
 { q(1-q^2)\chi_1(a)
 	\over 
 	(1-q)(1-q^{n+1})	
 }
 \bigg],
 \fe
 with enhanced $SU(2)$ flavor symmetry.
 
 Using the $q\to 1$ limit of the Lambert series, the Cardy limit can be easily obtained: 
 \ie
 \lim_{\tau \to 0} (-i \tau)\log \cI_{\rm Schur} ={\pi \over 6}.
 \fe
 Furthermore, taking into account the flavor fugacities $a=q^{i m}$, we obtain the 3d mass-deformed partition function:
 \ie
 Z_{3d}(m)=&{1\over (2\pi)^{n+1}}\prod_{s\in\{+1,-1\}}\Gamma\left({n+  2 s i m\over 2(n+1)}\right) \Gamma\left({n+2+  2 s im \over 2(n+1)}\right) \prod_{j=1}^{n-1} \Gamma\left({j\over n+1}\right)\Gamma\left({j+1\over n+1}\right)
 \\
 =&{1\over   2({n+1})}  {\sin \left( {\pi n\over  n+1 } \right) \over \sin \left( {\pi (n+ 2 i m)\over  n+1 } \right)\sin \left( {\pi (n- 2 i m)\over  n+1 } \right)}  .
 \label{A1Aodd4dZ}
 \fe
 By taking derivatives with respect to $m$, we obtain correlators of $J$ in the 1d TQM from the reduction of the 2d chiral algebra. In particular, two point function of the $U(1)$ moment map operator $J$ is
 \ie
 \la J \star J \ra= -{1\over (2\pi)^2 }\pa^2_m\left. \log Z_{3d}(m) \right|_{m=0}
 =-{1\over  (n+1)^2 \cos\left({\pi \over 2(n+1)}\right)^2}.
 \fe
 
 \subsubsection{Analysis in 3d}
 Here we derive the deformed sphere partition function directly from the 3d quiver gauge theory. For this purpose, we find it convenient to use the mirror description, in which case the FI parameters \eqref{A1AoddFI} translate into the mass parameters \eqref{A1AoddFIm} for the $SU(n)$ flavor symmetry of SQED, and the $U(1)$ mass parameter translates into the FI parameter of SQED,
\ie
 m= \eta.
 \fe
 In this case, using the identity \eqref{sumprodid}, we obtain:
 \ie
 Z_{3d}^{(A_1,A_{2n-1})}(m_a,\eta)
 =& {1\over (-i)^{n-1} (e^{\pi \eta}-(-1)^ne^{-\pi \eta})}\sum_{a=1}^n {e^{-2\pi i m_a \eta}\over \prod_{b\neq a} 2\sinh(\pi m_{ab})}
 \\
 =&{1\over   2(n+1)}  {\sin {\pi n\over n+1}\over \sin   {\pi (n+ 2 i \eta)\over2( n+1)} \sin   {\pi (n- 2 i \eta)\over2( n+1)}  },   
 \fe
 in agreement with \eqref{A1Aodd4dZ}. Correspondingly, the correlators of $J$ in the TQM (which can be obtained from taking  derivatives with respect to $\eta=m$) agree with the ones from reduction of the 4d Schur index.

 \subsection{$(A_1,D_{2n+2})$ AD theories}
 The 4d SCFT  has an $n$ complex dimensional Coulomb branch and a four complex dimensional Higgs branch, with $SU(2)\times U(1)$ global symmetry. The
 conformal central charges are
 \ie
 a_{4d}={12n+2\over 24},\quad c_{4d}={3n+1\over 6}.
 \fe
 This theory corresponds to $(A_{n+1}^{n+2}[-1], [n,1,1])$ in \cite{Wang:2015mra}. The Higgs branch of theory is given by the Slodowy slice of $\mf{sl}_{n+2}$ transverse to the sub-sub-regular nilpotent orbit $[n,1,1]$, and intersected with the subregular nilpotent orbit $[n+1,1]$, 
 \ie
 \cM_{\rm HB}=\overline{\mathbb{O}_{[n+1,1]}} \cap \cS_{[n,1,1]}.
 \fe
The chiral algebra is the qDS reduction of $V_{-{n(n+2)\over n+1}}(\mf{sl}_{n+2})$ at the subregular nilpotent element \cite{Song:2017oew}, 
 \ie
 \cW^{-{n(n+2)\over n+1}}(\mf{sl}_{n+2},[n,1,1]),
 \fe
which contains the AKM subalgebra $V_{-{2n+1\over n+1}}(\mf{sl}_2)$ responsible for the $SU(2)$ flavor symmetry of the 4d SCFT.  
 
 The 3d SCFT from $S^1$ reduction has the following quiver description,
 \ie
 &\xymatrix{
 	\boxed{SU(2)} \ar@{-}[r]_-{~Q_I,\tilde Q^I} & U(1) \ar@{-}[r]_-{q_1,\tilde q_1} & \cdots \ar@{-}[r]_-{q_{n-1},\tilde q_{n-1}} & U(1)\ar@{-}[r]_-{q_n,\tilde q_n} & \boxed{U(1)}\,,
 }
\label{A1Devenquiver}
 \fe
 with $n$ $U(1)$ gauge nodes. The mirror quiver is 
 \ie
 &\xymatrix{
 	\boxed{SU(n)} \ar@{-}[r]_-{~p_a,\tilde p_a} & U(1) \ar@{-}[r]_-{ s,\tilde s}  & U(1)\ar@{-}[r]_-{ t,\tilde t} & \boxed{U(1)}\,,
 }
\label{A1Devenmirrorquiver}
 \fe
 which makes manifest the enhanced $SU(n)\times U(1)$ symmetry on the 3d Coulomb branch. 
 
 The quiver gauge theory has hypermultiplets $(Q_{I},\tilde Q^{J})$ in the fundamental representation of $SU(2)$ (here $I,J=1,2$), and bifundamental hypermultiplets $(q_a,\tilde q_a)$ with $a=1,2,\dots, n$, which are
 subject to the D-term relations
 \ie
 Q_{I}\tilde Q^{I}=q_a\tilde q_a
 \fe
 for all $a$. 
 The Higgs branch chiral ring is generated by gauge invariant holomorphic combinations of hypermultiplets,  
 \ie
 M_{I}{}^J=Q_I \tilde Q^J, ~P=q_n\tilde q_n,~W_I=Q_I  q_1q_2\dots q_n
 ,~\tilde W^I=\tilde Q^I  \tilde q_1\tilde  q_2\dots \tilde q_n,
 \fe
 subject to the ring relations:
 \ie
 &M_1{}^1+M_2{}^2=P,\quad M_1{}^1 M_2{}^2=M_1{}^2 M_2{}^1,\quad
 \\
 &
 W_I\tilde W^J=M_I{}^J P^n,\quad M_I{}^J W_K=M_K{}^J W_I
 ,\quad M_I{}^J \tilde W^K=M_I{}^K\tilde  W^J.
 \fe
 
 The $(A_1,D_{2n+2})$ theory and the $(A_1,A_{2n-1})$ theory are related by Higgsing (as in the case of $(A_1,D_{2n+1})$ and $(A_1,A_{2n-2})$). This is obvious from the 3d perspective in which case the Higgsing is implemented by giving a vev to the moment map operator $M_I{}^J$. 
This is also easy to see in the class S setup using the $A_1$ $(2,0)$ theory. In this case, the $(A_1,D_{2n+2})$ is realized as $A_1^1[n]$, and the $(A_1,A_{2n-1})$ is realized as $(A_1^1[n],F)$, which involve the same irregular puncture \cite{Xie:2012hs,Wang:2015mra}. The two are obviously related by closing the regular $A_1$ puncture.
  The Coulomb branch chiral primary spectrum of the $(A_1,D_{2n+2})$ contains, in additional to those in $(A_1,A_{2n-2})$ (see \eqref{A1AoddCS}), an operator of dimension
 \ie
 \Delta = r={2n+1\over n+1},
 \label{A1DevenExtraOp}
 \fe
 which arises from local Hitchin moduli of the regular puncture and gets lifted upon Higgsing to $(A_1,A_{2n-2})$.
 
 Not surprisingly, the R-symmetry mixing relation that involves the 3d $SU(2)_C$ R-symmetry and emergent Coulomb branch topological symmetries is similar to that of $(A_1,A_{2n-1})$ \cite{Buican:2015hsa}. In fact, in the mirror description \eqref{A1Devenmirrorquiver}, the mixing is captured by the same mass deformations for the enhanced $SU(n)$ symmetry as in \eqref{A1AoddFIm} for SQED$_n$, which is the 3d mirror for the $(A_1,A_{2n-1})$ theory. Labeling the hypermultiplets in the mirror quiver as in \eqref{A1Devenmirrorquiver}, which makes manifest the SQED$_n$ subsector with hypermuliplets $p_a,\tilde p_a$, the identification between 4d Coulomb branch primaries and Higgs branch operators in the 3d mirror proceeds as before, with the exception of the operator \eqref{A1DevenExtraOp} which is identified to the dimension $3\over 2$ 3d chiral primary $\tilde p_1 \tilde s\tilde t$ (again up to mixing with operators of the same $U(1)_r$ quantum number).
 
 Using the mirror-map \eqref{A1DevenMM}, the corresponding FI parameters in the quiver \eqref{A1Devenquiver} are
 \ie
 \zeta_a={i\over \ell} \left({n-1\over 2n+2}, -{1\over n+1} ,\dots,-{1\over n+1} \right).
 \label{A1DevenFI}
 \fe
In the following sections, we verify that with these FI deformations, the reduction of the 4d flavored Schur index gives directly the 3d sphere partition function. Consequently, the subsector of the full TQM generated by the $U(1)$ moment map operator $J$ and the $SU(2)$ moment map operator $\cJ_I{}^J$ agrees with the Cardy limit of the corresponding sector in the W-algebra.

 \subsubsection{Analysis in 4d}
 The flavored Schur index for the $(A_1, D_{2n+2})$ theory is given by
 \ie
 \cI_{(A_1,D_{2n+2})}={\rm PE}\bigg[
 {(1+\chi_1(z))q(1-q^{n})+(a+a^{-1}) \chi_{1\over 2}(z) q^{n+1\over 2}(1-q)
 	\over 
 	(1-q)(1-q^{n+1})	
 }
 \bigg].
 \fe
 For $n=1$, we have  further simplification,
 \ie
 \cI_{(A_1,D_{4})}={\rm PE}\bigg[
 {(1+\chi_1(z))q+(a+a^{-1}) \chi_{1\over 2}(z) q
 	\over 
 	(1-q^{2})	
 }
 \bigg]
 ={\rm PE}\bigg[
 {  q \chi_{\bf 8}^{SU(3)}(y)
 	\over 
 	(1-q^{2})	
 }
 \bigg],
 \fe
with $SU(3)$ fugacities $y_1=z^{1/2} a$ and $y_2= z^{1/2} a^{-1}$,  due to the enhanced $SU(3)$ flavor symmetry.
 
 From the $q\to 1$ limit, on easily recovers the Cardy formula that accounts for the divergent pieces in $\log \cI_{(A_1,D_{2n+2})}$,
 \ie
 \lim_{\tau \to 0} (-i\tau )\log \cI_{(A_1,D_{2n+2})} =4\pi (c_{4d}-a_{4d})={\pi \over 3}.
 \fe
 Next defining $z=q^{m}$ and $a=q^{u}$, using the asymptotic behavior of the Lambert series, we extract the finite piece in the Cardy limit,
 \ie
 Z_{3d}(m)=&{1\over (2\pi)^{2(n+1)}}\prod_{s_1, s_2\in\{+1, -1\}}\Gamma\left({n+1+ 2s_1  u + 2s_2 m\over 2(n+1)}\right)  \prod_{j=1}^{n} \Gamma\left({j\over n+1}\right)^2 
 \Gamma\left({j+2   m \over n+1}\right) \Gamma\left({j-2   m \over n+1}\right) 
 \\
 =&{ 1\over  (n+1) 2^{n+2}}  {1\over  \cos {\pi (u+ m)\over n+1} \cos {\pi (u- m)\over n+1}  } \prod_{j=1}^{n} {1\over \sin {\pi (j+2 m)\over n+1}}.
 \label{A1Deven4dZ}
 \fe

 By taking derivatives with respect to $m$ and $u$ respectively, we compute the correlators in the 1d TQM from the $S^1$ reduction of the 2d chiral algebra,
 \ie
 \cJ_I{}^J\star \cJ_K{}^L=&\cJ_{IK}{}^{JL} 
 - {1\over 2\ell}(\D^J_K \cJ_I{}^L-\cJ_K{}^J \D_I^L)
 \\
 -& \frac{1}{12\ell^2} \left(\frac{1}{(n+1)^2}+2\right)(\D^L_I \D_K^J-{1\over 2}\D^J_I \D^L_K),
 \\
 J\star J =&:J^2:-{1\over 2(n+1)^2\ell^2} .
 \fe
 Here $\cJ_{IK}{}^{JL}$ denotes the normal ordered product in the OPE of $SU(2)$ moment-map operators, which is orthogonal to lower dimensional operators but has non-vanishing overlap with $:J^2:$ (similar to the case of $SU(3)$ SQCD discussed in Section~\ref{sec:LagrangianEx}).

 \subsubsection{Analysis in 3d}
 Here we directly compute the 3d sphere partition function with the FI parameters and mass deformations taken into account.
 
 Again, we find it convenient  to work with the  mirror description, in  which case the mass and FI parameters are given by (see \eqref{A1DevenMM} for the mirror-map)
 \ie
 m_a={n+1-2a\over 2(n+1)}i,\quad \eta_1=m-u,\quad \eta_2=2m.
 \fe
 In this case, the matrix model for the sphere partition function is given by
 \ie
 Z_{3d}^{(A_1,D_{2n+2})}(\{\eta_i\},\{m_a\})=&
 \int {d \sigma_1 d\sigma_2}{e^{2\pi i (\eta_1 \sigma_1+\eta_2\sigma_2)} \over 2\cosh (\pi \sigma_2) 2\cosh (\pi (\sigma_1-\sigma_2)) \prod_{a=1}^n 2\cosh (\pi (\sigma_1+m_a)}
 \\
 =&
 \int { d\sigma_2}{e^{2\pi i \eta_2\sigma_2} \over 2\cosh (\pi \sigma_2)}
 Z_{{\rm SQED}_{n+1}}(\eta_1,\{m_a,\sigma_2\}).
 \fe
  After some algebra (for details see Appendix~\ref{app:3dPFid}), we obtain:
  \ie
 & Z_{3d}^{(A_1,D_{2n+2})}(\{\eta_i\},\{m_a\})= 
% {1\over (-i)^{n} (e^{\pi \eta_1}+(-1)^ne^{-\pi \eta_1})}
% \\
% &\bigg(
% {i\over 2(n+1) \sinh \pi \eta_2} \left(
% {\sin {\pi n \over 2}\left(1+{2i \eta_1\over n+1}\right) \over \cosh {\pi \eta_1\over n+1}}
% -
% e^{\pi \eta_2}{\sin {\pi n \over 2}\left(1+{2i (\eta_1-\eta_2)\over n+1}\right) \over \cosh {\pi  (\eta_1-\eta_2)\over n+1}}
% \right)
% +
% {i^n\over 2(n+1)} {e^{n \pi (\eta_2-\eta_1) \over n+1} \over \cosh {\pi (\eta_2-\eta_1) \over n+1}}
% \bigg).
={1\over 4(n+1)}{\sinh {2\pi m\over n+1} \over   \cosh {\pi (m-u)\over n+1} \cosh {\pi (m+u)\over n+1}  \sinh {(2\pi m)  }}.
 \fe
This agrees with the answer \eqref{A1Deven4dZ} from reducing the 4d index after using the identity \eqref{sumprodid}. Consequently, all correlators of the $U(1)$ and $SU(2)$ moment map operators $J,\cJ_I{}^J$ agree with the Cardy limit of the W-algebra

\section{Conclusions and discussions}
 In this paper we have explored relations between protected operator algebras in four and three dimensional superconformal field theories (SCFT). They are respectively described by the 2d chiral algebra in 4d $\cN=2$ SCFTs and the 1d topological quantum mechanics (TQM) in 3d $\cN=4$ SCFTs. In particular, by taking the supersymmetric Cardy (high-temperature) limit of the 4d theory on $S^1\times S^3$ in a number of examples, we deduced explicit dictionaries between the 3d and 4d SCFTs for a subset of the operator product expansion (OPE) data, that includes  central charges for global symmetries and (twisted) correlation functions of Higgs branch BPS operators. We tested these relations by studying a variety of 4d SCFTs of both Lagrangian and non-Lagrangian type. In the former case, we saw explicitly  from supersymmetric localization that the 4d matrix model which captures observables in the corresponding chiral algebra sector becomes the 3d matrix model governing the TQM. In the latter non-Lagrangian case, despite the lack of localization formulae in 4d, we took proposals from the literature for the chiral algebra and in particular its torus partition function (the flavored Schur index of the 4d SCFT), and determined  a subset of the OPE data of the would-be 3d $\cN=4$ theory from the $S^1$ reduction. For some of these non-Lagrangian theories in 4d, such as the Deligne-Cvitanovi\'c (DC) exceptional series of SCFTs, and the $(A_1,A_{2n-1})$ and $(A_1,D_{2n+2})$ type Argyres-Douglas theories, the corresponding 3d (mirror) Lagrangians were known, and we used them to explicitly verify predictions from the 2d chiral algebra, therefore providing concrete support for these chiral algebra proposals.
 In other cases, such as the  $(A_1,A_{2n-2})$ and $(A_1,D_{2n+1})$ type Argyres-Douglas theories, where the 3d (mirror) Lagrangians were missing, we gave a proposal based on the Cardy limit of the 2d chiral algebra, that passes various nontrivial checks beyond  the chiral algebra/TQM subsector. Below we outline a number of interesting future directions:
 \begin{itemize}
 	\item For 4d non-Lagrangian SCFTs, a Higgs branch free field realization of the 2d chiral algebra sector was proposed in \cite{Beem:2019tfp}, which provides a systematic way to extract OPE data (similar in spirit to the localization formulae of \cite{Pan:2019bor,Dedushenko:2019yiw}). It would be interesting to explore the Cardy limit in this free field realization in relation to the TQM sector of the corresponding 3d SCFT.
 	\item
 	A related question is about the uplift of  matrix models describing the TQM sector of 3d $\cN=4$ SCFTs. Certainly, we do not expect all TQMs to be realized as reductions of 2d chiral algebras that arise in 4d $\cN=2$ SCFTs, due to the physical conditions such as superconformal symmetry and unitarity constraints in 4d. But it would be interesting to understand these obstructions of uplifting from the 3d/1d perspective.
 	\item 
 	We have mostly focused on local operator algebras in this paper. However, extended operators must be taken into account to establish a more complete dictionary between operator algebras across different spacetime dimensions. In particular, as we saw in the case of $(A_1,A_{2n-2})$ and $(A_1,D_{2n+1})$ theories, the Cardy limit generates emergent monopole operators of dimension $1/2$ and $1$ respectively, which do not originates from 4d local operators. They must come from line operators wrapping the vanishing $S^1$. It would be interesting to understand  this phenomena by studying the line operator spectrum of the 4d theory. More generally, the chiral algebra sector can be enriched by including surface defects, which leads to new observables in the 3d theory that may involve line and surface defects that are worth exploring. The TQM setup in 3d can independently be enriched by extended objects (\emph{e.g.}, a variety of line operators that 3d $\cN=4$ theories admit \cite{Dimofte:2019zzj}), which also remains mostly unexplored.
 	\item 
 	In all examples we considered in this paper, the 3d $\cN=4$ SCFT that arises in the Cardy limit of the 4d $\cN=2$ SCFT has a (mirror) Lagrangian description. It would be interesting to see if our general prescription can give predictions for the OPE data in non-Lagrangian 3d SCFTs using the input from the 2d chiral algebra sectors of 4d $\cN=2$ SCFTs. 

\item  For families of SCFTs with a large $N$ limit, via AdS/CFT, the protected operator algebras are dual to subsectors of string/M-theory in the bulk \cite{Bonetti:2016nma,Mezei:2017kmw,Mezei:2018url,Costello:2018zrm}. In particular, it was proposed in \cite{Bonetti:2016nma} that the 2d chiral algebra (a super W-algebra) sector of the 4d $\cN=4$ $SU(N)$ super-Yang-Mills in the large $N$ limit is described by a Chern-Simons theory on an AdS$_3$ slice, and the  gauge algebra is given by the wedge algebra of the large $N$ super W-algebra. Later in \cite{Mezei:2017kmw}, the 1d TQM sector of the large $N$ $U(N)_1\times U(N)_{-1}$ ABJM theory was shown to be dual to a 2d Yang-Mills theory on an AdS$_2$ slice in AdS$_4$. In this case the gauge algebra (in a $\bZ_2$ even sector) is given by ${\rm Sdiff}(S^2)$, the area-preserving diffeomorphisms on an emergent $S^2$, and the interactions involve, in addition to the Yang-Mills action, certain higher-derivative terms from the higher Casimir invariants of  ${\rm Sdiff}(S^2)$. It is plausible that the bulk dual 3d Chern-Simons and 2d Yang-Mills theories are related by an $S^1$ reduction (see for example  \cite{Aganagic:2004js}), and it would be interesting to explore this connection in more detail, especially since the Chern-Simons dual of the large $N$ super W-algebra is not completely settled.

\end{itemize}

 \section*{Acknowledgments}
We thank: T.~Creutzig, P.~Etingof, M.~Fluder, N.~Nekrasov for discussions. Part of this work was completed when M.D. was a member at (and supported by) the Walter Burke Institute for Theoretical Physics, with the additional support from the U.S. Department of Energy, Office of Science, Office of High Energy Physics, under Award No de-sc0011632, as well as the Sherman Fairchild Foundation.  The work of Y.W. is supported in part by the US NSF under Grant No. PHY-1620059 and by the Simons Foundation Grant No. 488653.

\appendix
\section{Computation using the 3d mirror of $(A_1, D_4)$}\label{app:ShiftOp}
The 3d mirror of the $(A_1, D_4)$ theory is a $U(1)\times U(1)$ gauge theory with hypers of charges $(1,0)$, $(1,1)$ and $(0,1)$. The Coulomb branch TQM of this is equivalentto the Higgs branch TQM of the direct dimensional reduction from 4d. We compute the Coulomb TQM here using the shift operator techniques from \cite{Dedushenko:2017avn,Dedushenko:2018icp}. 

Since we deal with the abelian rank-2 theory, all quantities of the type $\vec{X}$ will mean two-dimensional vectors. We use the North pole picture for the shift operators. We have the scalar fields,
\begin{equation}
\vec{\Phi}=\frac1{\ell} \left(\vec{\sigma} + \frac{i}{2}\vec{B} \right),
\end{equation}
and the monopoles,
\begin{equation}
\cM^{\vec b} = \left[\prod_{I=1}^3 \frac{(-1)^{(\vec{q}_I\cdot \vec{b})_+}}{\ell^{|\vec{q}_I\cdot \vec{b}|/2}}\left(\frac12 + i\ell\vec{q}_I\cdot \vec{\Phi} \right)_{(\vec{q}_I\cdot \vec{b})_+} \right] e^{-\vec{b}\cdot(\frac{i}2 \partial_{\vec{\sigma}}+\partial_{\vec B})},
\end{equation}
where $\vec{b}$ is the magnetic charge, and $\vec{q}_I$ is the gauge charge of the $I$'th hypermultiplet. These operators are the basic building blocks, and the algebra is simply generated by them, with all the relations following from the structure of shift operators.

We expect to obtain the quantization of the minimal nilpotent orbit of $A_2$, hence we use the typical Lie algebra notations below. We identify the following linear combinations of scalars with the Cartan subalgebra generators,
\begin{align}
H_1 &= \sqrt{2}\Phi_1 + \frac1{\sqrt 2}\Phi_2,\cr
H_2 &= \sqrt{\frac32}\Phi_2,
\end{align}
and the following monopoles with the $A_2$ root system,
\begin{align}
\cM^{(\pm 1,0)} &= E^{\pm \alpha_1},\cr
i\cM^{(\mp1,\pm1)} &= E^{\pm \alpha_2},\cr
\cM^{(0,\pm1)} &= E^{\pm (\alpha_1+\alpha_2)},
\end{align}
where $\alpha_1$ and $\alpha_2$ are the simple roots of $A_2$.

These operators generate the whole algebra of shift operators, and in particular we can easily check that commutators indeed reproduce the $A_2$ algebra,
\begin{align}
[H_i, E^\alpha] &= \frac{i}{\ell} \alpha_i E^\alpha,\cr
[E^\alpha, E^{-\alpha}] &= \frac{i}{\ell} \alpha\cdot H,\cr
[E^{\alpha_2}, E^{\alpha_1}] &= \frac{i}{\ell} E^{\alpha_1+\alpha_2},\cr
&\text{\emph{etc.}}
\end{align}
The normalization of the Killing form is
\begin{equation}
K(E^\alpha, E^{-\alpha})= K(H_i, H_i)=6,
\end{equation}
which implies that $\psi^2=2$ in our conventions. We then compute the Casimir element,
\begin{equation}
C = (H_1^2 + H_2^2) + \sum_{\alpha\in\Delta}E^\alpha \star E^{-\alpha} = \frac{6}{4\ell^2}.
\end{equation}
To bring the algebra into our canonical form, we identify
\begin{equation}
\hbar=\frac{i}{\ell},
\end{equation}
which then implies the same value of $\mu$ as the Higgs computation in the main text,
\begin{equation}
\mu=-\frac38.
\end{equation}

\section{One point functions of current bilinears}\label{app:JJ_1pf}
Here we compute the $\tau\to0$ asymptotics of the torus one point functions of current bilinears $(J_i^j J_k^l)$ and $(JJ)$. This can be done using the explicit representation of the chiral algebra of Lagrangian theories on $S^3\times S^1$ by symplectic bosons coupled to the matrix model, as derived in \cite{Dedushenko:2019yiw,Pan:2019bor}. There, one first computes correlators at the fixed value of $a$ -- the $S^1$ holonomy which is the matrix model variable appearing in \eqref{matrmodZ}. Then one integrates over $a$. Using the torus propagator for symplectic bosons (see eqn. (4.49) in \cite{Dedushenko:2019yiw}), and the expressions of currents in terms of symplectic bosons, it straightforward to find the following,
\begin{align}
&\langle J_i^j(z) J_k^l(0)\rangle_a=\left(\delta_i^l\delta_k^j - \frac1{N_f}\delta_i^j\delta_k^l \right)\frac{-1}{(2\pi)^6\ell^2}\left[\frac{\theta_1'(0;\tau)}{\theta_1\left(\frac{z}{2\pi i};\tau\right)}\right]^2\trace_\cR\left[\frac{\theta_4\left(\frac{z}{2\pi i}-\frac{a}{2\pi};\tau\right)}{\theta_4\left(-\frac{a}{2\pi};\tau\right)}\frac{\theta_4\left(-\frac{z}{2\pi i}-\frac{a}{2\pi};\tau\right)}{\theta_4\left(-\frac{a}{2\pi};\tau\right)}\right]\cr
&=-\left(\delta_i^l\delta_k^j - \frac1{N_f}\delta_i^j\delta_k^l \right)\frac{1}{(2\pi)^6\ell^2}\left[\frac{2\pi i}{z}-\frac16 \frac{\theta_1'''(0;\tau)}{\theta_1'(0;\tau)}\frac{z}{2\pi i} + O(z^3) \right]^2\times\cr
&\times\trace_\cR\left[ 1 + \left(\frac{z}{2\pi i} \right)^2\left(\frac{\theta_4''(\frac{a}{2\pi};\tau)}{\theta_4(\frac{a}{2\pi}; \tau)}-\left(\frac{\theta_4'(\frac{a}{2\pi};\tau)}{\theta_4(\frac{a}{2\pi}; \tau)}\right)^2\right)+O(z^4)\right]\cr
&=-\left(\delta_i^l\delta_k^j - \frac1{N_f}\delta_i^j\delta_k^l \right)\frac{1}{(2\pi)^6\ell^2}\trace_\cR\left\{-\frac13 \frac{\theta_1'''(0;\tau)}{\theta_1'(0;\tau)} + \frac{\theta_4''(\frac{a}{2\pi};\tau)}{\theta_4(\frac{a}{2\pi}; \tau)}-\left(\frac{\theta_4'(\frac{a}{2\pi};\tau)}{\theta_4(\frac{a}{2\pi}; \tau)}\right)^2 \right\}\cr
&+{\rm singular} + O(z^2),
\end{align}
where the final expression gives $\langle (J_i^j J_k^l)\rangle_a$. In the similar computation for $J$,
\begin{align}
\langle J(z)J(0)\rangle_a &= \frac{N_f^2}{(2\pi)^6\ell^2} \left( \trace_\cR\left[ \frac{\theta_4'\left(\frac{a}{2\pi};\tau\right)}{\theta_4\left(\frac{a}{2\pi};\tau\right)}\right] \right)^2 \cr
&-\frac{N_f}{(2\pi)^6\ell^2}\left[\frac{\theta_1'(0;\tau)}{\theta_1\left(\frac{z}{2\pi i};\tau\right)}\right]^2\trace_\cR\left[\frac{\theta_4\left(\frac{z}{2\pi i}-\frac{a}{2\pi};\tau\right)}{\theta_4\left(-\frac{a}{2\pi};\tau\right)}\frac{\theta_4\left(-\frac{z}{2\pi i}-\frac{a}{2\pi};\tau\right)}{\theta_4\left(-\frac{a}{2\pi};\tau\right)}\right],\cr
\end{align}
the first term comes from self-contractions in $J$. Subtracting the pole, this give:
\begin{align}
\langle(JJ)\rangle_a &= \frac{N_f^2}{(2\pi)^6\ell^2} \left( \trace_\cR\left[ \frac{\theta_4'\left(\frac{a}{2\pi};\tau\right)}{\theta_4\left(\frac{a}{2\pi};\tau\right)}\right] \right)^2\cr 
&-\frac{N_f}{(2\pi)^6\ell^2}\trace_\cR\left\{-\frac13 \frac{\theta_1'''(0;\tau)}{\theta_1'(0;\tau)} + \frac{\theta_4''(\frac{a}{2\pi};\tau)}{\theta_4(\frac{a}{2\pi}; \tau)}-\left(\frac{\theta_4'(\frac{a}{2\pi};\tau)}{\theta_4(\frac{a}{2\pi}; \tau)}\right)^2  \right\}.
\end{align}
It is easier to first compute the $\tau\to 0$ behavior, and then perform the matrix model integration, which also simplifies in this limit, as we saw in \eqref{matri_mod_red}. We use the following,
\begin{align}
\frac{\theta_1'''(0;\tau)}{\theta_1'(0;\tau)} &= -\frac{\pi^2}{\tau^2}-\frac{6\pi i}{\tau} + O(\tilde{q})\,,\cr
\frac{\theta_4'(z;\tau)}{\theta_4(z;\tau)} &= -\frac{2\pi iz}{\tau} -\frac{\pi}{\tau}\tan(\pi z/\tau) + O(\tilde{q})\,,\cr
\frac{\theta_4''(z;\tau)}{\theta_4(z;\tau)}&= -\left( \frac{2\pi z}{\tau} \right)^2 + \frac{(2\pi)^2iz}{\tau^2}\tan(\pi z/\tau) - \frac{\pi^2}{\tau^2} -\frac{2\pi i}{\tau} + O(\tilde{q})\,,
\end{align}
and find that
\begin{align}
\lim_{\tau\to 0}\left( \tau^2 \langle(J_i^j J_k^l)\rangle_a \right) &= -\frac{1}{\pi^2(8\pi\ell)^2}\left(\delta_i^l\delta_k^j - \frac1{N_f}\delta_i^j\delta_k^l \right)\trace_\cR\left\{\frac13 - \frac1{\cosh^2\pi\sigma} \right\}\,,\cr
\lim_{\tau\to 0}\left( \tau^2 \langle(JJ)\rangle_a \right) &= -\frac{N_f^2}{\pi^2(8\pi\ell)^2}\left( \trace_\cR \tanh \pi\sigma \right)^2 - \frac{N_f}{\pi^2(8\pi\ell)^2}\trace_\cR\left\{\frac13 - \frac1{\cosh^2\pi\sigma} \right\}\,.\cr
\end{align}
These expressions completely agree with the ones that previously entered the computation of $\mu_1, \mu_2$ in 3d (c.f. \eqref{JJ_from_3d}), once we take the linear relation \eqref{AB_from_mu} between $A, B$ and $\mu_1, \mu_2$ into account. Therefore, we see a complete agreement between the purely 3d computation and the 4d$\to$3d reduction of the VOA sector.

\section{3d sphere partition functions and identities}\label{app:3dPFid}

The sphere partition function of a general 3d $\cN=4$ quiver gauge theory $\cT$ with gauge group $G\equiv \otimes_{i=1}^n (G_i)$ and Higgs branch flavor symmetry $G_H$ reduces to a matrix model over the real Coulomb branch scalar vevs, by performing a supersymmetric localization computation.\footnote{For a comprehensive review on the sphere partition functions of 3d supersymmetric gauge theories see \cite{Willett:2016adv}.}

For notational simplicity, we define
\ie
\ch(x)\equiv 2\cosh (\pi x),\quad \sh (x)\equiv 2\sinh (\pi x).
\fe
We also set $\ell=1$ in this appendix.

The matrix model for the quiver gauge theory $\cT$ then takes the form \footnote{It is possible to include Chern-Simons levels for the gauge group, but we will not need them here.}
\ie
Z_{\cT}(\{\eta_i\},\{m_a\})={1\over |W|}\int \prod_{\rm cartan} d\sigma e^{2\pi i \sum_{i=1}^k \eta_i \tr_i \sigma}
{\det_{\rm ad} \sh (\sigma)
	\over  
	\det_{{\rm R}  \otimes {\rm S} } \ch (\sigma+m)
},
\fe
where ${\rm R}  \otimes {\rm S}$ labels the (reducible) representation of the hypermultiplets under $G\otimes G_H$. We have included the FI parameters $\eta_i$ and mass parameters $m_a$ that are compatible with the localization procedure. The mirror symmetry for 3d $\cN=4$ gauge theories amounts to the relation  
\ie
Z_{\cT}(\{\eta_i\},\{m_a\})=Z^{\rm mirror}_{\cT}(\{\xi_a\},\{u_i\}),
\fe
with certain identification (or the mirror-map) between the mass and FI parameters $(\{\eta_i\},\{m_a\})$ and $(\{\xi_a\},\{u_i\})$ of the mirror-dual pair.

For example, the sphere partition function for SQED$_n$ with FI and mass deformations is 
\ie
Z_{{\rm SQED}_n}(\eta,\{m_a\})
=&
\int d\sigma {e^{2\pi i \sigma \eta}\over 
	\prod_{a=1}^n  \ch (\sigma+m_a)
}
\\
=& {1\over (-i)^{n} (e^{\pi \eta}+(-1)^ne^{-\pi \eta})}
\sum_{a=1}^n {e^{-2\pi i m_a \eta}\over   \prod_{b\neq a} \sh (m_{ab})},
\label{SQEDpf}
\fe
where the integral was evaluated explicitly in \cite{Benvenuti:2011ga}.

On the other hand, the mirror description involves a linear quiver \eqref{A1Aoddquiver} with $n-1$ $U(1)$ gauge group nodes,
\ie
Z_{{\rm SQED}_n}^{\rm mirror}(\{\xi_a\},u)
=
\int  \prod_{a=1}^{n} {d \sigma_a }{e^{2\pi i \sum_{a=1}^{n-1} \xi_a \sigma_a}\over \ch(\sigma_1)\ch(\sigma_n+u)\prod_{a=1}^{n-2}\ch (\sigma_a-\sigma_{a+1}) } .
\fe
Using the Fourier transformation repeatedly,
\ie
{1\over \ch(x)}=\int d x {e^{2\pi i x y} \over \ch (y)},
\fe
it is straightforward to derive the identification:
\ie
{}&Z_{{\rm SQED}_n}(\eta,\{m_a\})=Z_{{\rm SQED}_n}^{\rm mirror}(\{\xi_a\},u),
\\
&{\rm Mirror~map:~~}
\begin{cases}
	\xi_a=m_a-m_{a-1},\\
	u=\eta .
\end{cases}
\label{A1AoddMM}
\fe

For the quiver \eqref{A1Devenquiver} that arises from the $S^1$ reduction of the $(A_1,D_{2n+2})$ theory, the matrix model is given by
\ie
Z_{(A_1,D_{2n+2})}(\{\xi_a\},\{u_1,u_2\})=&\int \prod_{a=1}^n d\sigma_a e^{2\pi i \sigma_a \xi_a} { 1 \over \ch(\sigma_1 \pm u_1) \ch(\sigma_n+u_2) \prod_{a=1}^{n-1} \ch(\sigma_a-\sigma_{a+1}) }  ,
\fe
and similarly for the mirror quiver \eqref{A1Devenmirrorquiver},
\ie
Z_{(A_1,D_{2n+2})}^{\rm mirror}(\{\eta_1,\eta_2\},\{m_a\})=&\int {d \sigma_1 d\sigma_2 }{e^{2\pi i (\sigma_1 \eta_1+\sigma_2\eta_2)}\over \ch(\sigma_1-\sigma_2) \ch(\sigma_2) \prod_{a=1}^n \ch(\sigma_1+m_a) }.
\fe
Once again, using Fourier transform, we derive the identification
\ie
{}&Z_{(A_1,D_{2n+2})}^{\rm mirror}(\{\eta_1,\eta_2\},\{m_a\})=Z_{(A_1,D_{2n+2})}(\{\xi_a\},\{u_1,u_2\}),
\\
&{\rm Mirror~map:~~}
\begin{cases}
	(\xi_1,\xi_2,\dots, \xi_n)=(m_1,m_2-m_1,\dots,m_n-m_{n-1}),\\
	(u_1,u_2)=({\eta_2-2\eta_1\over 2},{\eta_2\over 2}).
\end{cases}
\label{A1DevenMM}
\fe 

We determine the explicit form of $Z_{(A_1,D_{2n+2})}(\{\eta_1,\eta_2\},\{m_a\})$ below for the special case,
\ie
m_a={n+1-2a\over 2(n+1)}i,
\fe
relevant for the reduction of the 4d SCFT on $S^1$. We start by rewriting the two dimensional Coulomb branch integral as
\ie
Z_{(A_1,D_{2n+2})}^{\rm mirror}(\{\eta_1,\eta_2\},\{m_a\}) 
=&
\int { d\sigma_2}{e^{2\pi i \eta_2\sigma_2} \over \ch (\sigma_2)}
Z_{{\rm SQED}_{n+1}}(\eta_1,\{m_a,\sigma_2\}).
\fe
Using \eqref{SQEDpf}, we have 
\ie
&Z_{{\rm SQED}_{n+1}}(\eta,\{m_a,\sigma_2\})
\\
=& {1\over (-i)^{n} (e^{\pi \eta}+(-1)^ne^{-\pi \eta})}
\bigg(
\sum_{a=1}^n {e^{-2\pi i m_a \eta}\over 2\sinh(\pi (m_{a}-\sigma_2)) \prod_{b\neq a} 2\sinh(\pi m_{ab})}
+
{e^{-2\pi i \sigma_2 \eta}\over \prod_{b=1}^n 2\sinh(\pi (\sigma_2-m_b))}
\bigg).
\fe
Therefore,
\ie
&Z_{(A_1,D_{2n+2})}^{\rm mirror}(\{\eta_1,\eta_2\},\{m_a\}) 
\\
=&
{1\over (-i)^{n} (e^{\pi \eta_1}+(-1)^ne^{-\pi \eta_1})}
\bigg(
\sum_{a=1}^n {-i e^{-2\pi i m_a \eta_1}\over  \prod_{b\neq a} 2\sinh(\pi m_{ab})}
Z_{{\rm SQED}_{2}}( \eta_2,\{0,i/2-m_a\})
+
i^nZ^{n+1}_{\rm QED}(\eta_{21},\{0,i/2-m_b\})
\bigg)
\\
\equiv &
{1\over (-i)^{n} (e^{\pi \eta_1}+(-1)^ne^{-\pi \eta_1})}
(I_1+I_2).
\fe

Using \eqref{SQEDpf} again, we obtain
\ie
I_1
=&
{i\over 2(n+1) \sinh \pi \eta_2} \left(
{\sin {\pi n \over 2}\left(1+{2i \eta_1\over n+1}\right) \over \cosh {\pi \eta_1\over n+1}}
-
e^{\pi \eta_2}{\sin {\pi n \over 2}\left(1+{2i (\eta_1-\eta_2)\over n+1}\right) \over \cosh {\pi  (\eta_1-\eta_2)\over n+1}}
\right),
\\
I_2 =& {i^n\over 2(n+1)} {e^{n \pi (\eta_2-\eta_1) \over n+1} \over \cosh {\pi (\eta_2-\eta_1) \over n+1}}.
\fe
Putting together the two terms above and simplifying, we obtain
\ie
&Z_{(A_1,D_{2n+2})}^{\rm mirror}(\{\eta_1,\eta_2\},\{m_a\}) 
\\
=&{1\over 4(n+1)}{\sinh {2\pi m\over n+1} \over   \cosh {\pi (m-u)\over n+1} \cosh {\pi (m+u)\over n+1}  \sinh {(2\pi m)  }}.
\fe
We find useful the following product and sum identities involving trigonometric functions,
\ie
{}&\prod_{a=1}^n \sin   {\pi   (a+2im )\over 2 (n+1)} ={\sinh (2\pi m) \over \sinh \left( {2\pi m\over n+1}\right)},
\\
&\sum_{a=1}^n {e^{{ \pi(n+1-2a)\over n+1}\eta}  \sin  { \pi  a\over n+1}  }
=
{ \cosh \pi \eta \sin {\pi \over n+1}
	\over 
	\sin {\pi (1\pm 2i\eta )\over 2(n+1)}
}.
\label{sumprodid}
\fe
The first equality above implies in particular 
\ie
{}&\prod_{a=1}^n \sin   {\pi   a\over n+1}= \prod_{a=1}^n \cos   {\pi   (n+1-2a)\over 2 (n+1)}  ={n+1\over 2^n}.
\fe

\bibliographystyle{utphys}
\bibliography{VOA_and_DQ}

\end{document}